\documentclass[twocolumn]{aastex63}

\shorttitle{\mgii\ in the solar chromosphere}

\usepackage{natbib}
\newcommand{\mgii}{\ion{Mg}{2}} 
\newcommand{\kone}{k$_1$}
\newcommand{\ktwo}{k$_2$} 
\newcommand{\ktwov}{k$_{2v}$}
\newcommand{\ktwor}{k$_{2r}$}
\newcommand{\kthree}{k$_3$}

\newcommand{\feix}{\ion{Fe}{9}}
\newcommand{\htime}[2]{{#1}~{\rm h}~{#2}~{\rm m}}
\newcommand{\nanow}{nW~m$^{-2}$~Hz$^{-1}$~sr$^{-1}$}

\newcommand{\iris}{{\em IRIS}}

\newcommand{\uBz}{\langle|B_z|\rangle}

\newcommand{\longacknowledgment}{We gratefully acknowledge support by NASA grants 80NSSC20K1272, and contract NNG09FA40C (IRIS). The simulations have been run on the Pleiades cluster through the computing project s1061, and s2278 from the High End Computing (HEC) division of NASA and on the Betzy cluster through computing project nn2834k from the Norwegian Sigma2 High-Performance Computing center. This research is also supported by the Research Council of Norway through its Centres of Excellence scheme, project number 262622.}

\begin{document}

\title{Numerical simulations and observations of \mgii\ in the solar chromosphere}

\author[0000-0003-0975-6659]{Viggo H. Hansteen}
\affil{Lockheed Martin Solar \& Astrophysics Laboratory,
3251 Hanover St, Palo Alto, CA 94304, USA}
\affil{Bay Area Environmental Research Institute,
NASA Research Park, Moffett Field, CA 94035, USA.}
\affil{Rosseland Center for Solar Physics, University of Oslo, P.O. Box 1029 Blindern, N-0315 Oslo, Norway}
\affil{Institute of Theoretical Astrophysics, University of Oslo,
P.O. Box 1029 Blindern, N-0315 Oslo, Norway}

\author[0000-0002-0333-5717]{Juan Martinez-Sykora}
\affil{Lockheed Martin Solar \& Astrophysics Laboratory,
3251 Hanover St, Palo Alto, CA 94304, USA}
\affil{Bay Area Environmental Research Institute,
NASA Research Park, Moffett Field, CA 94035, USA.}
\affil{Rosseland Center for Solar Physics, University of Oslo, P.O. Box 1029 Blindern, N-0315 Oslo, Norway}
\affil{Institute of Theoretical Astrophysics, University of Oslo,
P.O. Box 1029 Blindern, N-0315 Oslo, Norway}

\author[0000-0001-9218-3139]{Mats Carlsson}
\affil{Rosseland Center for Solar Physics, University of Oslo, P.O. Box 1029 Blindern, N-0315 Oslo, Norway}
\affil{Institute of Theoretical Astrophysics, University of Oslo,
P.O. Box 1029 Blindern, N-0315 Oslo, Norway}

\author[0000-0002-8370-952X]{Bart De Pontieu}
\affil{Lockheed Martin Solar \& Astrophysics Laboratory,
3251 Hanover St, Palo Alto, CA 94304, USA}
\affil{Rosseland Center for Solar Physics, University of Oslo, P.O. Box 1029 Blindern, N-0315 Oslo, Norway}
\affil{Institute of Theoretical Astrophysics, University of Oslo,
P.O. Box 1029 Blindern, N-0315 Oslo, Norway}

\author[0000-0002-5879-4371]{Milan Go\v{s}i\'{c}}
\affil{Lockheed Martin Solar \& Astrophysics Laboratory,
3251 Hanover St, Palo Alto, CA 94304, USA}
\affil{Bay Area Environmental Research Institute,
NASA Research Park, Moffett Field, CA 94035, USA.}

\author[0000-0002-2180-1013]{Souvik Bose}
\affil{Lockheed Martin Solar \& Astrophysics Laboratory,
3251 Hanover St, Palo Alto, CA 94304, USA}
\affil{Bay Area Environmental Research Institute,
NASA Research Park, Moffett Field, CA 94035, USA.}
\affil{Rosseland Center for Solar Physics, University of Oslo, P.O. Box 1029 Blindern, N-0315 Oslo, Norway}
\affil{Institute of Theoretical Astrophysics, University of Oslo,
P.O. Box 1029 Blindern, N-0315 Oslo, Norway}

\begin{abstract}
The \mgii~h\&k lines are amongst the best diagnostic tools of the upper solar chromosphere. This region of the atmosphere is of particular interest as it is the 
lowest region of the Sun's atmosphere where the magnetic field is dominant in the energetics and dynamics, defining its structure. While highly successful in the photosphere and lower to mid chromosphere, numerical models have produced synthetic \mgii\ lines that do not match the observations well. We present a number of large scale models with magnetic field topologies representative  of the quiet Sun, ephemeral flux regions and plage, and also models where the numerical resolution is high and where we go  beyond the MHD paradigm. The results of this study show models with a much improved correspondence with \iris\ observations both in terms of intensities and widths, especially underscoring the importance of chromospheric mass loading and of capturing the magnetic field topology and evolution in simulations. This comes in addition to the importance of capturing the generation of small scale velocity fields and including non-equilibrium ionization and ion-neutral interaction effects. Understanding and modeling all these effects and their relative importance is necessary in order to reproduce observed spectral features. 
\end{abstract}


\section{Introduction}~\label{sec:intro}

The solar chromosphere is the region where the energetics are controlled by a non-thermal ``mechanical'' heating which becomes dominant in setting the density, magnetic field, and temperature structure. In the lower chromosphere, especially in the quiet Sun, this mechanical heating, driven by granular convective motions, consists of primarily acoustic processes \citep{1997ApJ...481..500C}, and 3D MHD simulations are capable of reproducing the most salient aspects of the spectral lines formed there \citep[e.g.,][and references cited within]{2009ApJ...694L.128L}. 

However, 
even while these advanced models reproduce many features of the chromosphere, they also show significant discrepancies for lines whose cores form above the $\beta=1$ ($\beta\equiv p_g/p_B$ where $p_g$ is the gas pressure and $p_B$ the magnetic pressure $B^2/2\mu_0$) layer, some 750~km above the photosphere in the quiet Sun. Clearly, the magnetic field plays a vital role at greater heights, and this role is not fully understood. This issue has become particularly pressing when interpreting the vast amounts of data collected by the \iris\ satellite \citep{2014SoPh..289.2733D}, especially in the \mgii\ lines. 

\mgii\ lines are uniquely sensitive to the middle and upper chromospheric conditions and cannot be obtained from ground-based observatories. In radiative MHD Bifrost simulations, synthetic \mgii\ lines often
came out too faint,  strongly asymmetric, or too narrow \citep{2019ARAA..57..189C}. These discrepancies could have many causes; such as indicating a lack of opacity, lack of heating, a lack of small-scale motions in the chromosphere, and/or 3D effects on the radiative transfer \citep{Judge2020b}. 

\begin{figure}   \includegraphics[width=0.46\textwidth]{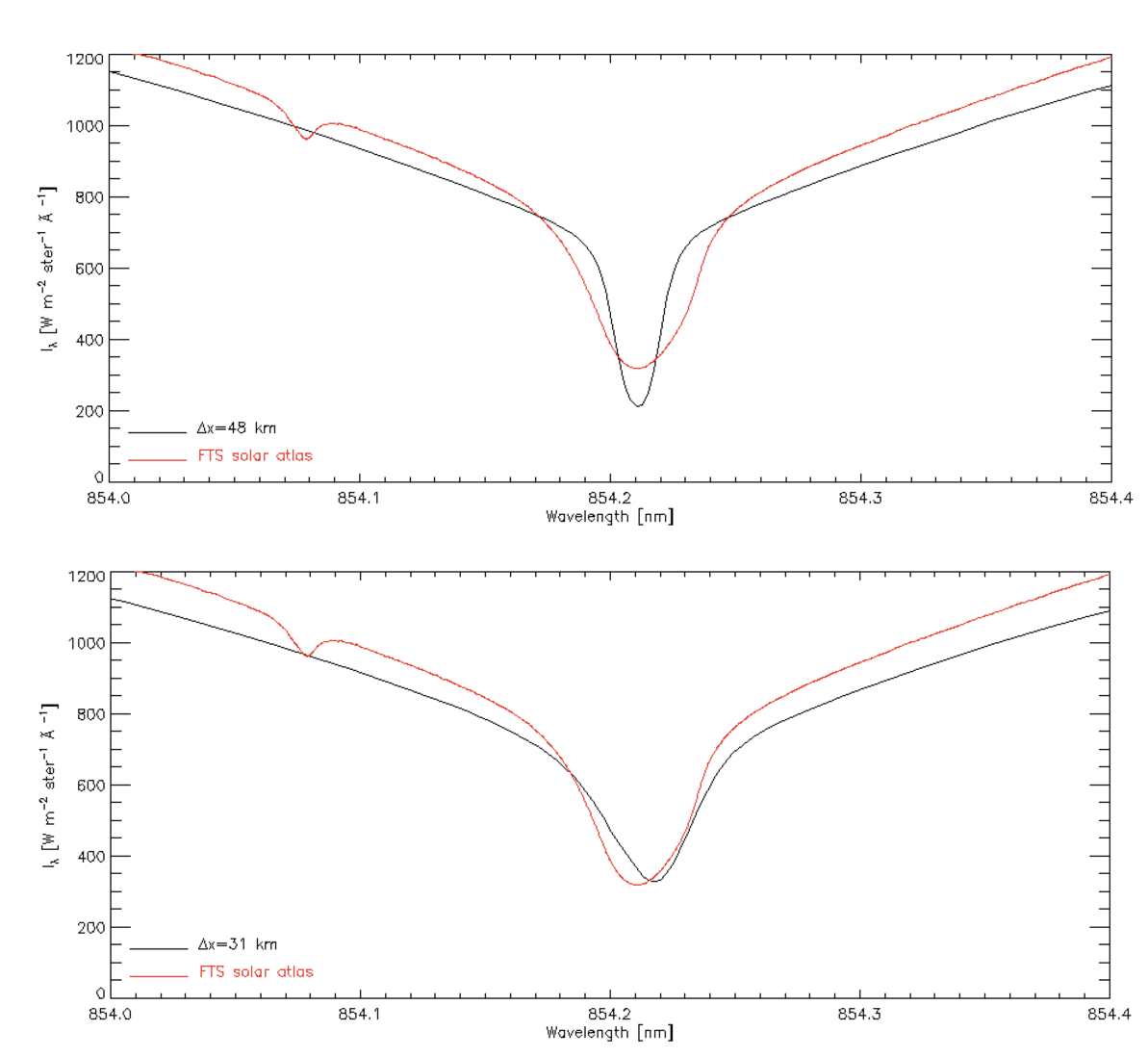}
	\caption{\label{fig:caii_resolution} The \ion{Ca}{2}~854.2~nm line at two different resolutions, 48~km and 31~km horizontal grid size, as computed from Bifrost simulations. The observed mean solar profile \citep{1987Brault_Neckel} is plotted in red.}
\end{figure}

The lack of spatial resolution in radiative MHD simulations will cause a reduction in the production of small scale motions or turbulence, and hence result in narrower synthetic line profiles than those observed. As an example, consider the \ion{Ca}{2}~854.2~nm profile discussed in \citet{2009ApJ...694L.128L} which was computed with a horizontal resolution of 64~km. This original calculation was redone with a resolution of 48~km, which also produced a profile significantly narrower than that observed. However, when the simulation was repeated with a spatial resolution of 31~km, the correspondence between the line core widths was found to be much better. This  indicates both that vigorous large scale dynamics and turbulent motions at smaller scales could be reproduced in higher resolution models and that for this mid-chromospheric line, it is the turbulent motions that determine the core width. This can be seen in Figure~\ref{fig:caii_resolution} (Mats Carlsson, private communication) which shows the \ion{Ca}{2}~854.2~nm profile at 48~km and 31~km horizontal resolution along with the observed profile taken from the FTS Solar Atlas \citep{1987Brault_Neckel}.  This also sets an upper limit on the amplitude of turbulent motions in the lower to mid chromosphere and implies that a numerical resolution of, say, 20--30~km is good enough to capture relevant solar dynamics in this region. 

Effects that go beyond the standard MHD description of the chromospheric plasma, including, e.g., generalized Ohm's Law \citep[GOL, e.g. ][]{2018A&A...618A..87K}, including ambipolar diffusion and non-equilibrium (NEQ) ionization of H and He will also alter the opacity of the \mgii\ lines, by altering the heating and electron density profiles of the chromosphere and thus potentially the intensity and width of the emergent lines \citep{Martinez-Sykora:2017gol,2021A&A...654A..51B,2022A&A...664A..91P}. There are other smaller-scale candidate physical process for heating the chromosphere as well: the Thermal-Farley-Buneman instability \citep[TFBI,][]{2020ApJ...891L...9O}. Preliminary multi-fluid simulations of the TFBI show a temperature increase due heating \citep{2022arXiv221103644E}. TFBI also provides a large increase of the turbulent motions during the non-linear phase of the instability which could broaden chromospheric spectral lines. 

Furthermore, the lack of chromospheric heating and dynamics in 3D MHD models could be caused by not properly capturing the magnetic topology of the outer solar atmosphere correctly. The magnetic field has been an almost free parameter of 3D models, and at small spatial scales not well constrained by observations. This problem has been partially mitigated by recent models of photospheric, near surface, simulations \citep{2014ApJ...789..132R,2018ApJ...859..161R} showing the importance and depth dependence of the local dynamo in generating quiet Sun-like fields. The topology and evolution of the magnetic field impacts the structure of the chromosphere both through the heating rate and through the Lorentz force which can carry and support material in excess of hydrostatic equilibrium, changing the density, temperature structure and opacity, especially in the $\beta < 1$ region of the mid- to upper chromosphere (as well as in the corona). Flux emergence is one agent which will change all of these parameters as it can bring plasma up to chromospheric heights where it can reside long after the emergence phase is over. However, it is still unclear how much small scale flux emergence occurs regularly outside of newly forming active regions. This is true for both mature active regions and plage, where the rates are not known. Some progress has been made for typical quiet Sun regions, where \citet{2022ApJ...925..188G} measure 68~Mx~cm$^{-2}$~day$^{-1}$, perhaps as a result of a local dynamo.

The problem of correctly reproducing \mgii\ intensities and core widths is particularly pressing when chromospheric plage is considered. A series of ``semi-empirical'' models  \citep{2015ApJ...809L..30C} was constructed in order to look into what effect varying chromospheric parameters had on the \mgii\ profile in plage-like conditions. The goal of this was to find how a chromospheric atmosphere can produce ``single peaked'' profiles, where the k$_2$ peaks and the k$_3$ minimum share nearly the same intensity and where the line core width is as wide as observed. The paper gives a number of parameters that conspire to change the \mgii\ profile; the electron density $n_{\rm e}$, the temperature $T_{\rm g}$, and the turbulent velocity $v_{\rm turb}$. Furthermore, \citet{2015ApJ...809L..30C}
found that the total intensity of the \mgii\ profile is dependent on the temperature, or equivalently pressure, of the overlying corona, which sets the column mass at which the transition region temperature rise occurs and hence the density of the upper chromosphere.

Inversions of observed \iris\ spectra tell essentially the same story: In order to reproduce the observed \mgii\ profiles the (plage) chromosphere needs to be dense, extended, and hot, perhaps also with a large (5--6 km/s) turbulent velocity in the region of line core formation \citep{2020A&A...634A..56D}

In this paper we continue these studies by comparing synthetic profiles of \mgii\ for a set of numerical simulations of varying spatial resolution, magnetic topology, and physics considered to \iris~observations of quiet Sun, active region and plage. 

\section{Overview of IRIS observations} \label{sec:iris_obs}

We have chosen four large \iris~raster scans to serve as a basis of comparison between simulated and observed \mgii\ spectra. Additionally, we present HMI-magnetograms \citep{2012SoPh..275..207S,2012SoPh..275..327S} of the same regions to present maps of the magnetic environment. The size of the regions chosen are roughly the same as the simulation box presented later in this paper, hence they should in principle host magnetic field topologies of roughly the same scale. The raster scans cover two typical quiet Sun regions, one long-lived apparently unipolar plage region, and an active region sunspot with surrounding plage. We find a range of average \mgii\ profile intensities, shapes, and widths in sub-regions covering of order $10\times10$\arcsec. 

\begin{figure*}
    \includegraphics[width=0.98\textwidth]{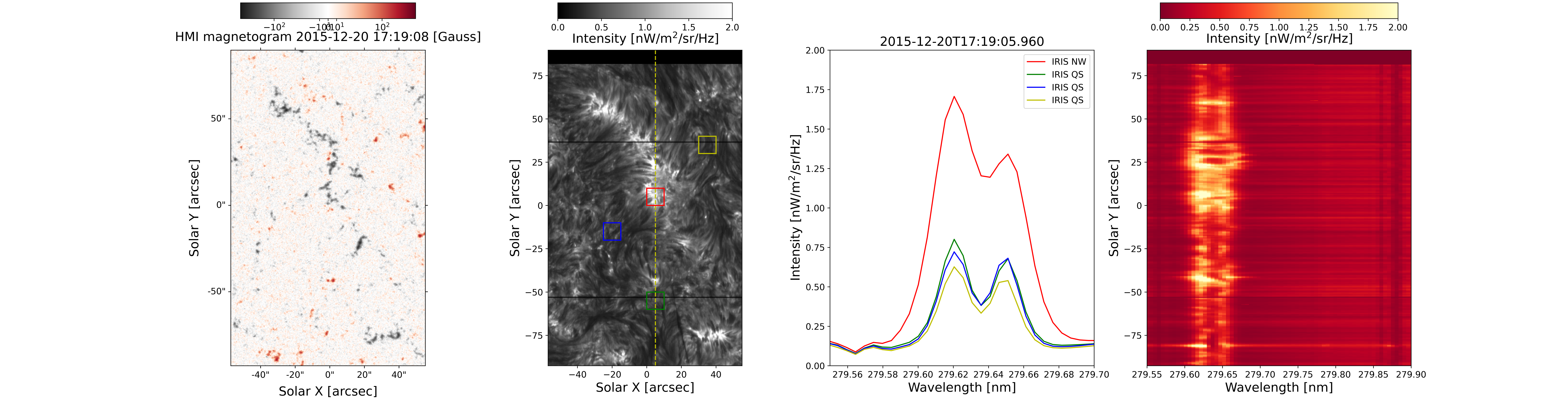}
    \includegraphics[width=0.98\textwidth]{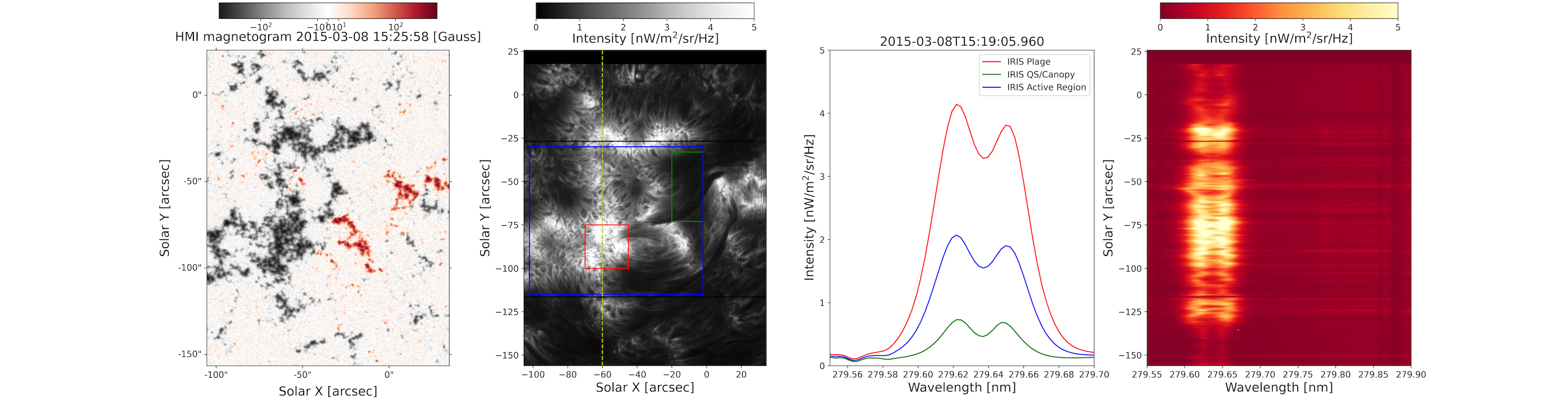}
    \includegraphics[width=0.98\textwidth]{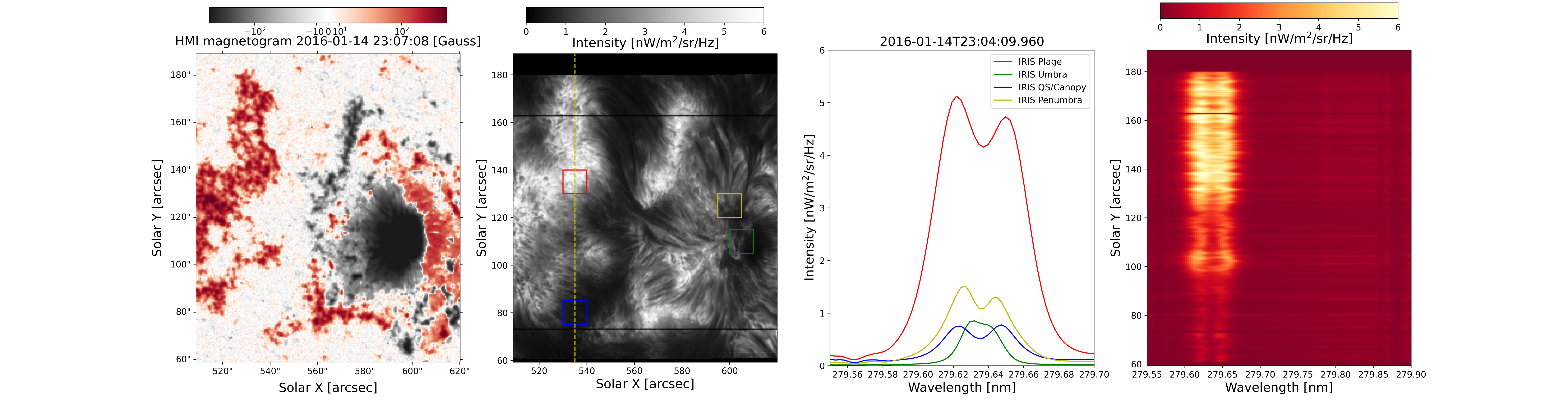}
	\caption{\label{fig:iris_obs} Three examples of Quiet Sun, Plage/Active Region (NOAA 12296), and Active Region (NOAA 12480). The left panels show HMI magnetograms of the field of view, scaled to $\pm 750$ Gauss, while the remaining panels show the \mgii\ k line as observed with IRIS. Regions of interest are delineated by red, green, blue,and in the cases of the Quiet Sun and AR 12480 panels, also yellow, boxes.}
\end{figure*}

The upper row of Figure~\ref{fig:iris_obs} shows a ``typical'' quiet Sun location. When considering quiet Sun profiles in the following we note the difference, but largely refer to both (NW) and internetwork (IN) profiles as quiet Sun (QS). When averaging we implicitly assume that the spatial filling factor of NW vs. IN is the same in the observations and the simulations.  To the right of the co-temporal HMI magnetograms, we show the \mgii\ k-line spectra taken at Sun center on 2015 December 20. The magnetogram shows weak magnetic fields across the entire image with small patches of stronger ($>100$~Gauss) fields of both polarities and a diagonal band of strong negative polarity field stretching from solar $(x,y)\approx(-40,30)$\arcsec\ to $(x,y)\approx(20,-30)$\arcsec. The average \mgii\ profile is double peaked with a fairly deep \kthree~core --- half the strength of the \ktwo~ peaks for the darker examples, somewhat less deep for the network (`IRIS NW') box in red. The network patch also shows an average profile that has a slight asymmetry, with the violet \ktwo~peak some 30\% stronger than the red peak. The \ktwo/\kthree~intensities are all of order 1~\nanow, or less, for internetwork regions with FWHM widths measured to $0.051$~nm (corresponding to roughly 50~km~s$^{-1}$), while the network emission is stronger and wider with intensities of 1.7~\nanow\ and width of $0.056$~nm. The second quiet Sun raster scan was taken on 2017 October 15 and previously analyzed by \citet{Martinez-Sykora2022arXiv221015150M}. We refer to that paper for further details. As shown later, the peak separation and intensities of this observation resemble the quiet Sun regions (though not the network) of the third panel on the top row of Figure~\ref{fig:iris_obs}. 

In the middle row the central portions of NOAA 12296, on 2015 March 8, are shown to consist mainly of negative polarity plage, with some positive polarity plage near $(x,y)~(-25,-75)$\arcsec\ stretching towards the NorthEast. The positive polarity plage is actually mostly remnants of the NOAA 12192 active region, one of the largest sunspot groups of Cycle 24, which crossed the central meridian five rotations earlier, on 2014 October 23. In 2015 March, the negative polarity plage has dispersed and stretches from the equator to -500\arcsec\ S. The small bipole forming NOAA 12296 emerged into this band of negative polarity near the equator at least a week before the crossing of the central meridian and appears to be fading, but a new AR, NOAA 12298, forms in the same location four days later (2015 March 12), so there may be weak flux emergence occurring continuously near this location. The quiet Sun (or canopy) box chosen lies between negative and positive polarities, and the average \mgii\ profile in this box is very similar to the quiet Sun profiles shown in the upper row, although it is a factor of two brighter. We note that what we call QS/canopy may be significantly affected by the neighboring active region plage and associated canopy, which is why we refer to these profiles as ``QS/canopy" rather than just ``QS". The blue box, outlining the total extent of the AR, has a higher intensity with the \ktwo/\kthree\ peaks at intensities of 2~\nanow\ and with width $0.059$~nm. The red box outlines an area of strong plage emission with maximum \ktwo/\kthree\ intensities of more than 4~\nanow\ and FWHM width of $0.059$~nm, similar to the entire AR.

The lower row shows data taken on 2016 January 14, just after 23:00~UT, focusing on the NOAA 12480 active region. This AR appeared on the East limb on 2016 January 6 and was fully emerged at that point. 
The IRIS raster covers both a positive polarity plage region to the East of a large sunspot, and between these polarities less bright quiet Sun/chromospheric canopy structures. The average plage profile (red) is very similar to that seen in the NOAA 12296 raster, shown in the middle row, with intensities of more than $5$~\nanow\ and width of $0.059$~nm. Likewise, the quiet Sun spectra are similar to the quiet Sun cases discussed above; here, in blue, with intensities of $0.8$~\nanow\ and FWHM width of $0.048$~nm. The yellow box covering the penumbra shows intensities of $1.5$~\nanow\ and width $0.041$~nm. Finally, the umbral spectra in green are amongst the narrowest found with FWHM width of $0.025$~nm and relatively low intensity $0.85$~\nanow. Note that the umbra is on average single peaked with no clear \kthree\ minimum.

\section{Models} \label{sec:models}

We have run several simulations using the Bifrost MHD code \citep{2011AA...531A.154G} to model the photosphere and outer solar atmosphere.  In this work we consider models with different resolutions, physics and field topology. The extension of physics 
includes non-equilibrium hydrogen and/or helium ionization \citep{2007AA...473..625L,2016ApJ...817..125G} and/or generalized Ohm's law \citep[GOL,][]{2020A&A...638A..79N}. 

In its base configuration Bifrost solves the equations of MHD using an energy equation that includes optically thick radiative transfer 
\citep{2010AA...517A..49H}, a tabulated form of effectively and optically thin radiative losses \cite{2012AA...539A..39C}, and Spitzer thermal conductivity along magnetic field lines. The latter is solved either implicitly via operator splitting and utilizing a multi-grid solver, or explicitly along with the MHD equations using a hyperbolic formulation, which limits the speed of conduction fronts allowing a reasonable time step, as described by \citet{2017ApJ...834...10R}.
The equation of state is, for the ``simple'' models, based on table lookups of the LTE ionisation state of a plasma of solar abundance.

In short, three of the models presented here are run with relatively coarse (100~km) horizontal resolution and include only ``simple'' MHD physics extending over a $72\times 72\times 60$~Mm box. The convection zone is modeled to a depth of 8.5~Mm, and the models reach more than 50~Mm above the photosphere. The size of this computational box allows the capture of granular to supergranular (or at least mesogranular) size scales. 


Other horizontally extended domains are the three simulations already described in detail in \citet{Martinez-Sykora:2017gol} and \citet{Martinez-Sykora:2020ApJ...889...95M}. These simulations are 2.5D and span $90\times42.8$~Mm box. The model contains two plage-like regions connected with $\sim40$~Mm long loops and reveals features resembling type I and II spicules (the latter only with the presence of ambipolar diffusion), low-lying loops, and other physical processes. The simulation spans a vertical domain stretching from $\sim3$~Mm below the photosphere to $40$~Mm above into the corona with a non-uniform vertical grid size of $12$~km in the photosphere and chromosphere and $14$~km grid size in the horizontal axis. The three models differ in the included physics, one with only the ``simple'' Bifrost configuration (called {\tt spicule nGOL} in the following), one with GOL ({\tt spicule GOL}), and the third with non-equilibrium ionization of hydrogen and helium and GOL ({\tt spicule GOL NEQ(H,He)}). 

The other two models considered cover small domains ($6\times 6\times 10$~Mm) but with very high resolution (5~km horizontally) of an internetwork field as detailed in \citep{2019ApJ...878...40M}.
The two simulations differ in the physics included; one without GOL or non-equilibrium ionization {called \tt QS nGOL} in the following), the other with non-equilibrium ionization of hydrogen and GOL ({\tt QS GOL, NEQ(H)}).
The simulation spans a vertical extent stretching from $\sim2.5$~Mm below the photosphere to $8$~Mm above. A non-uniform vertical grid is employed with a size of $4$~km in the photosphere and chromosphere, somewhat larger outside these regions. Initially, the simulation box is seeded with a uniform weak vertical magnetic field of $2.5$~G. From this starting point a local dynamo is active and generated a magnetic field that reaches a statistically steady state with $B_{\rm rms} = 57$~G at photospheric heights \citep[similar to that described by][]{2007AA...465L..43V,2014ApJ...789..132R,2015Sci...347.1333C}. These models also generated an in-situ magnetic field in the chromospheric portion of their domain. 

In all models, the upper boundary is based on a characteristic extrapolation of the variables, which in principle allows waves to exit the computational box without reflection \citep[see][for details]{2011AA...531A.154G}. The temperature gradient, and hence conductive flux in the vertical direction, is set to zero so that no heat enters the box from above. At the lower boundary, the entropy of the material flowing into the computational box at the bottom boundary is set so that the effective temperature is close to solar. 

In order to produce synthetic diagnostic profiles of the optically thick \mgii\ lines (as well as the lines of \ion{Ca}{2}) we employ the RH1.5D code \citep{2001ApJ...557..389U,2015ascl.soft02001P}. This code performs multi-level non-local thermodynamical equilibrium calculations with partial frequency redistribution. RH calculates spectra from 3D atmospheric models on a column-by-column basis. Note that the 1.5D nature of the solution process will miss some of the effects of horizontal transfer \citep[see ][where 3D effects on \mgii\ are discussed]{2013ApJ...772...90L,2017A&A...597A..46S}, which we do not believe are vital for the analysis performed in this paper, though see also \citet{Judge2020b} for a slightly alternate view. 

\begin{figure*}
    \includegraphics[width=0.49\textwidth]{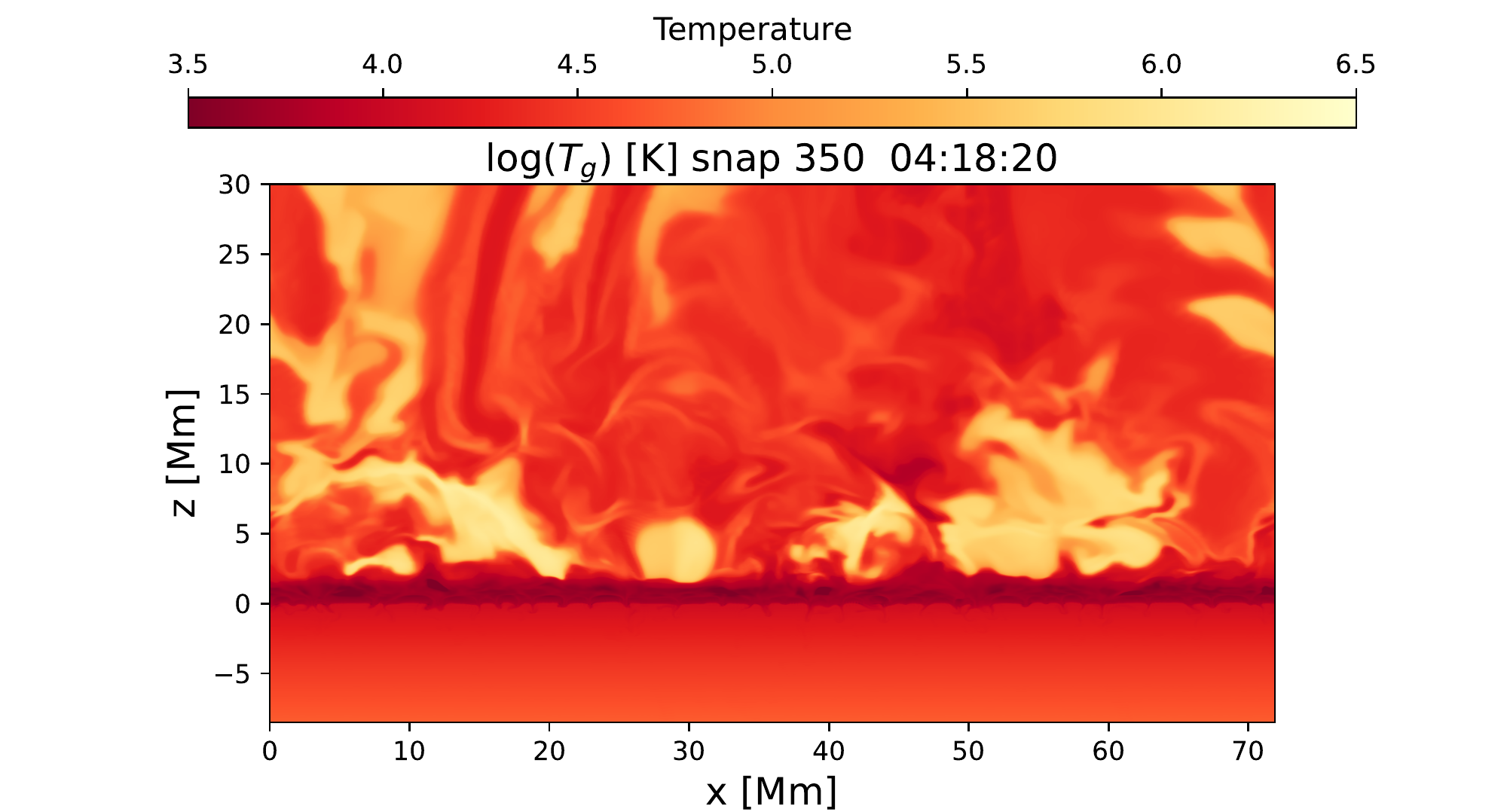}
    \includegraphics[width=0.49\textwidth]{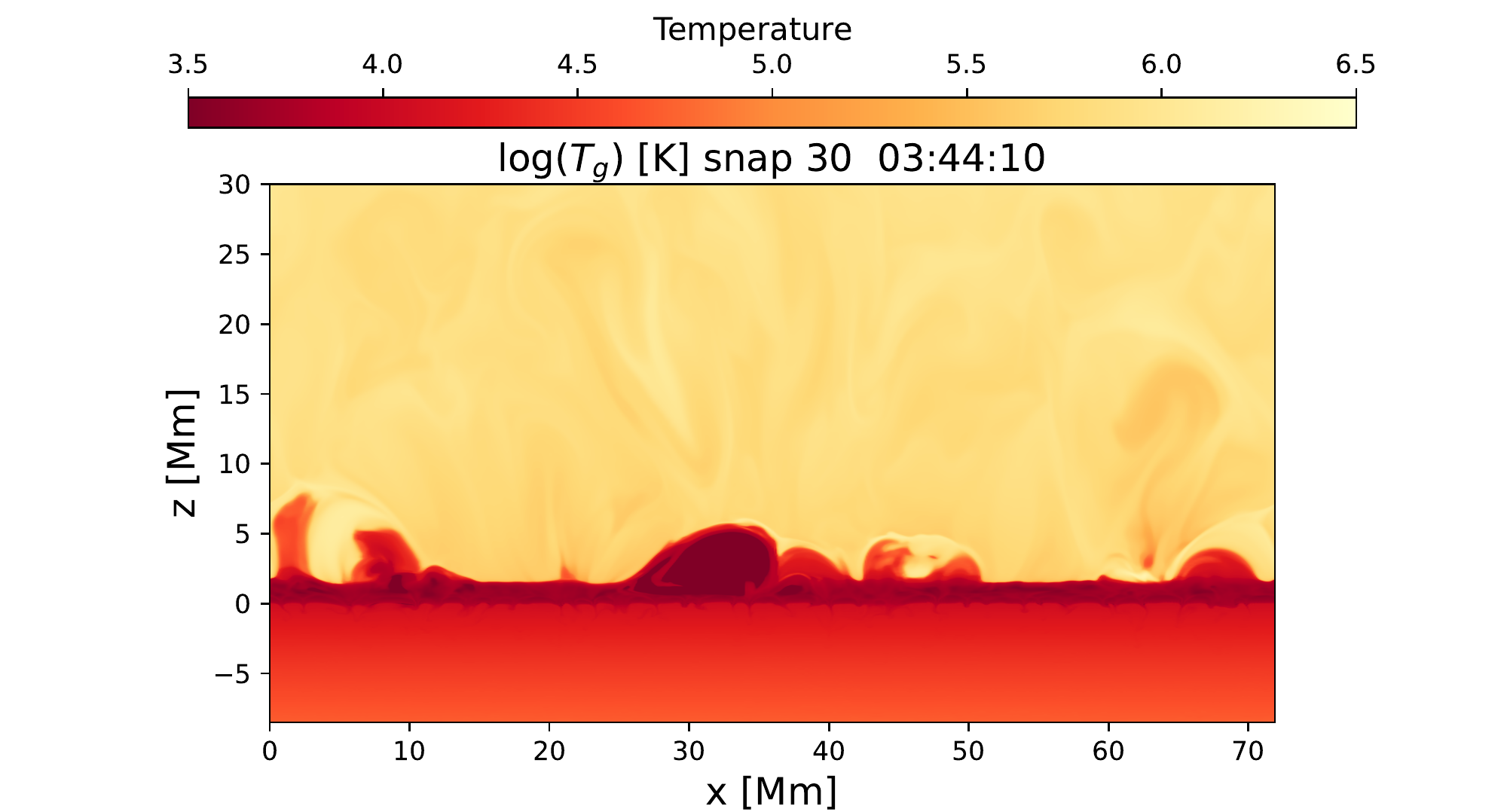}
	\caption{\label{fig:qs_tgstruct} Temperature structure at $y=32$~Mm in the {\tt nw072100} and {\tt qs072100} models. }
\end{figure*}

\begin{figure*}
    \includegraphics[width=0.98\textwidth]{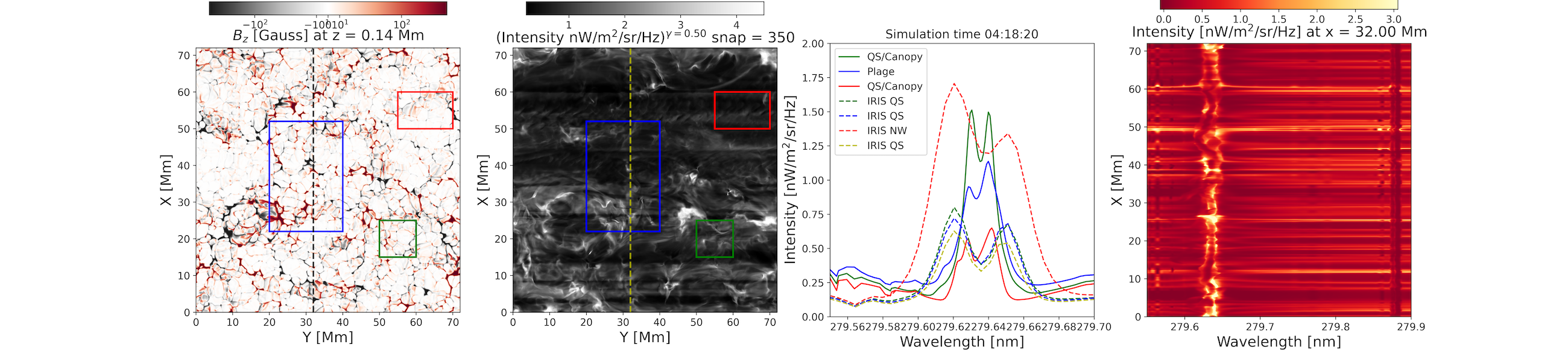}
    \includegraphics[width=0.98\textwidth]{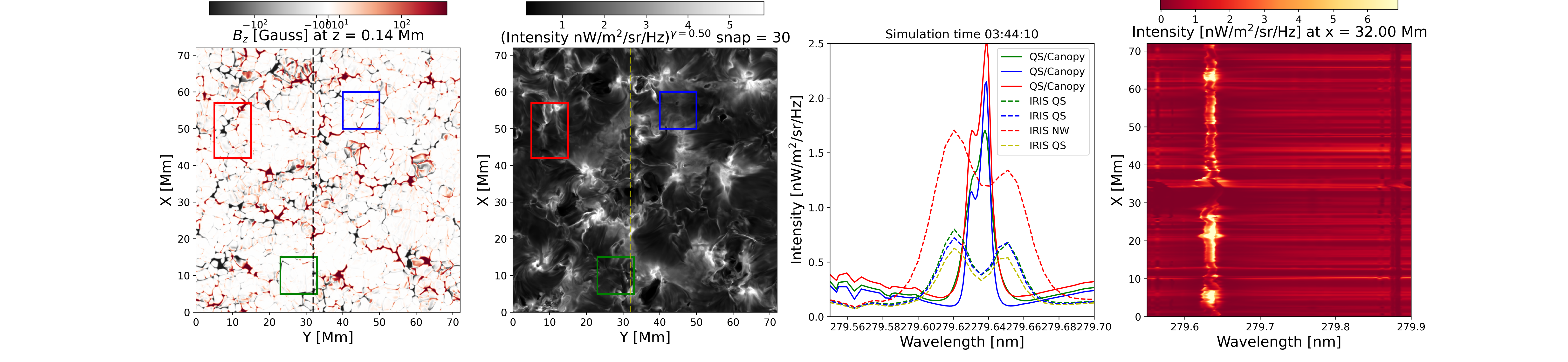}
	\caption{\label{fig:qs_profiles} ``Quiet Sun'' model profiles. The layout is the same as in Figure~\ref{fig:iris_obs} while, for comparison with the model, we have added the Quiet Sun profiles from the top row of that figure as dashed lines.}
\end{figure*}

\subsection{Quiet Sun model(s)} \label{sed:quiet_sun_models}

Let us begin our analysis by considering low resolution, 3D MHD, ``quiet Sun'' models (with the relationship between the quiet Sun and these models yet to be clarified). These large scale models were initialized with a thermodynamically relaxed convection zone, photosphere and lower chromosphere. Two different models were derived from this starting point: one, called {\tt nw072100} henceforth, with an average horizontal field of 100~Gauss in the convection zone and photosphere, as well as a very weak horizontal field at coronal heights. In the second, called {\tt qs072100} in the following, a vertical field of 5~Gauss is included in addition to the horizontal field used in {\tt nw072100}. In {\tt nw072100}, a strong horizontal flux sheet is imposed at the lower boundary, but has not yet reached the photosphere for the profiles discussed in this section. Both models were allowed to evolve for several hours solar time, and a salt and pepper magnetic field is rapidly established in the photosphere with a mean unsigned $B_z$ of 30~Gauss (mean $B$ 60~Gauss) which is smaller than the observed quiet Sun field of $\langle |B_z|\rangle=60$~Gauss \citep[e.g.][]{2012ApJ...751....2O} by roughly a factor of two.

Though the photospheric field strengths in these models are very similar, we find quite different coronal temperature structures as shown in Figure~\ref{fig:qs_tgstruct}, presumably due to the differing magnetic field topologies: The {\tt nw072100} model, which initially had a nearly horizontal field, originally achieved very high temperatures as portions of the initial 100~Gauss photospheric field expands into the corona. This hot corona took several hours to cool, since thermal conduction along the magnetic field could not form an effective cooling mechanism when the connection to the transition region and chromosphere is tenuous. However, the model cools eventually and finally reached a minimum temperature state after some 3~hours solar time (a vertical slice of the temperature structure of the model is shown in Figure~\ref{fig:qs_tgstruct}). Spatially highly variable the average temperature lies between 200~kK and 300~kK from the top of the chromosphere, at 2~Mm, up to 15~Mm above the photosphere. The average temperature decreases to some 100~kK above this height.  After this time the average coronal temperature rises slowly, before rising rapidly when new flux emerges from the photosphere as described in section~\ref{sec:emerging_flux}. 

The {\tt qs072100} model was also initially hot, but with a significant amount of vertical field, it cools much more efficiently. The vertical field component also allows the spread of high temperatures more easily from localized braiding-caused heating events. Thus, the temperature structure appears much smoother than that found in {\tt nw072100}, as shown in Figure~\ref{fig:qs_tgstruct}. Of particular interest to this paper are the chromospheres of both models, both of which are of order 2~Mm in vertical extent, but with elements reaching up to 5~Mm above the photosphere, as seen in the figure. 

Figure~\ref{fig:qs_profiles} shows the vertical photospheric magnetic field $B_z$, the \mgii\ line core intensity, and the \mgii\ line profiles. As mentioned above, the photospheric field forms a salt-and-pepper like pattern, organized on a larger scale into what appear to be network cells with diameters of order 10--20~Mm. The fields are concentrated into small patches where we find field strengths up to $\pm 2000$~Gauss, while as mentioned above, the average vertical field in both models is of order $\uBz=30$~Gauss at the time the figure represents. The distribution of the magnetic field is similar in both models, though we do see one clear instance of flux emergence in the {\tt nw072100} model near $(x,y)\sim ([10-20],[20-30])$. The \mgii\ \kthree\ line core image reflects this network pattern to a certain extent: intensities are high in the same locations that the field is strong. 

We also see some remnants of the horizontal field that initially filled the corona as weak horizontal stripes aligned with the $y$-axis where cool gas remains trapped in the corona. The amount of cool gas at great heights may be greater than what one finds in the typical quiet Sun outside of prominences or regions of strong flux emergence, while interesting, the effect this has on the \mgii\ profile appears to be small, aside from a small broadening that disappears in the hours following this snap shot before flux emerges and the corona reheats.

We have chosen three boxes with dimensions of order $10\times10$~Mm in which we compute the average profiles from both quiet Sun models\footnote{We have defined ``quiet Sun'' in the models by choosing patches which remain relatively free of strong large scale fields for the duration of the simulation. The relation between these patches and and network and internetwork regions on the Sun remains to be determined.}. 
The observed quiet Sun \ktwo\ and \kthree\ intensities have radiation temperatures of roughly $T_{\rm rad}=5000$~K, with the network brighter at $T_{\rm rad}=5500$~K. The model shown in the upper row, {\tt nw072100}, displays radiation temperatures as low as 4900~K and up to 5500~K for internetwork and network intensities respectively. On the other hand, the model shown in the bottom row, {\tt qs072100} is somewhat hotter, with \ktwo, \kthree\ intensities equivalent to radiation temperatures between 5300~K and 5600~K. The {\tt qs072100} coronal temperature has stabilized, but we note that the coronal density at 10~Mm is (still) decreasing at the time of this snap shot, having fallen from $4\times 10^{9}$~cm$^{-3}$ to $2.5\times 10^{9}$~cm$^{-3}$ in the half hour preceding the displayed snap shot. The difference between these models' average chromospheres (below 2 Mm) is slight; both have an average Joule heating of 0.3~W~m$^{-3}$ at $z=1.0$~Mm above the photosphere, but the {\tt qs072100} model, shown in the lower row, has a slightly hotter corona with an average temperature of 600~kK at $z=10$~Mm, while the {\tt nw072100} model, shown in the top row has an average of only 250~kK at $z=10$~Mm at this time. We will discuss how coronal temperatures can impact the \mgii\ intensities further in Section~\ref{sec:magi_formation} but note that the coronal temperature plays an important role in setting the \mgii\ intensity level. This point is also made by \citet{2015ApJ...809L..30C} and recently by \citet{Bose2022} who note the correspondence between bright \mgii\ core emission and sites of high intensity \feix~17.1~nm in moss regions.  

\begin{figure}
    \includegraphics[width=0.49\textwidth]{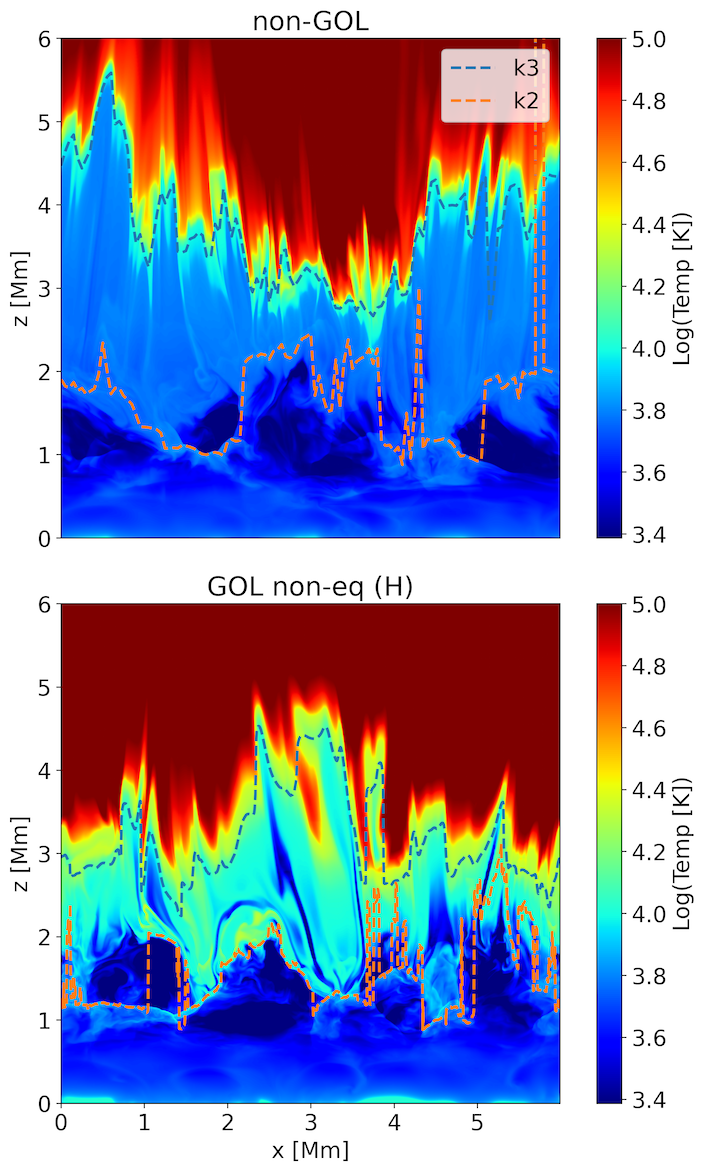}  
    \caption{\label{fig:qs_tgstruct_gol} Temperature structure of high resolution (4~km) quiet Sun models run both with standard MHD (non-GOL, top) and with generalized Ohm's law (GOL, bottom) including non-equilibrium hydrogen ionization. The heights of the $\tau_\nu=1$ layers for the \ktwo\ and \kthree\ spectral locations are shown as dashed lines.}
\end{figure}

While the synthetic quiet Sun \mgii\ intensities lie well within the range of observed values, some interesting differences are also clear. Profiles in the cooling {\tt qs072100} model show stronger \ktwor\ peaks than \ktwov, while the {\tt nw072100} profiles, while still showing some asymmetry, are more balanced. More importantly, the \mgii\ core widths are seen to be much narrower than what is observed. In the regions covered in both models of Figure~\ref{fig:qs_profiles} varying between 0.025~nm and 0.03~nm FWHM which is roughly half of the observed quiet Sun and network profiles. 


\begin{figure}
     \includegraphics[width=0.43\textwidth]{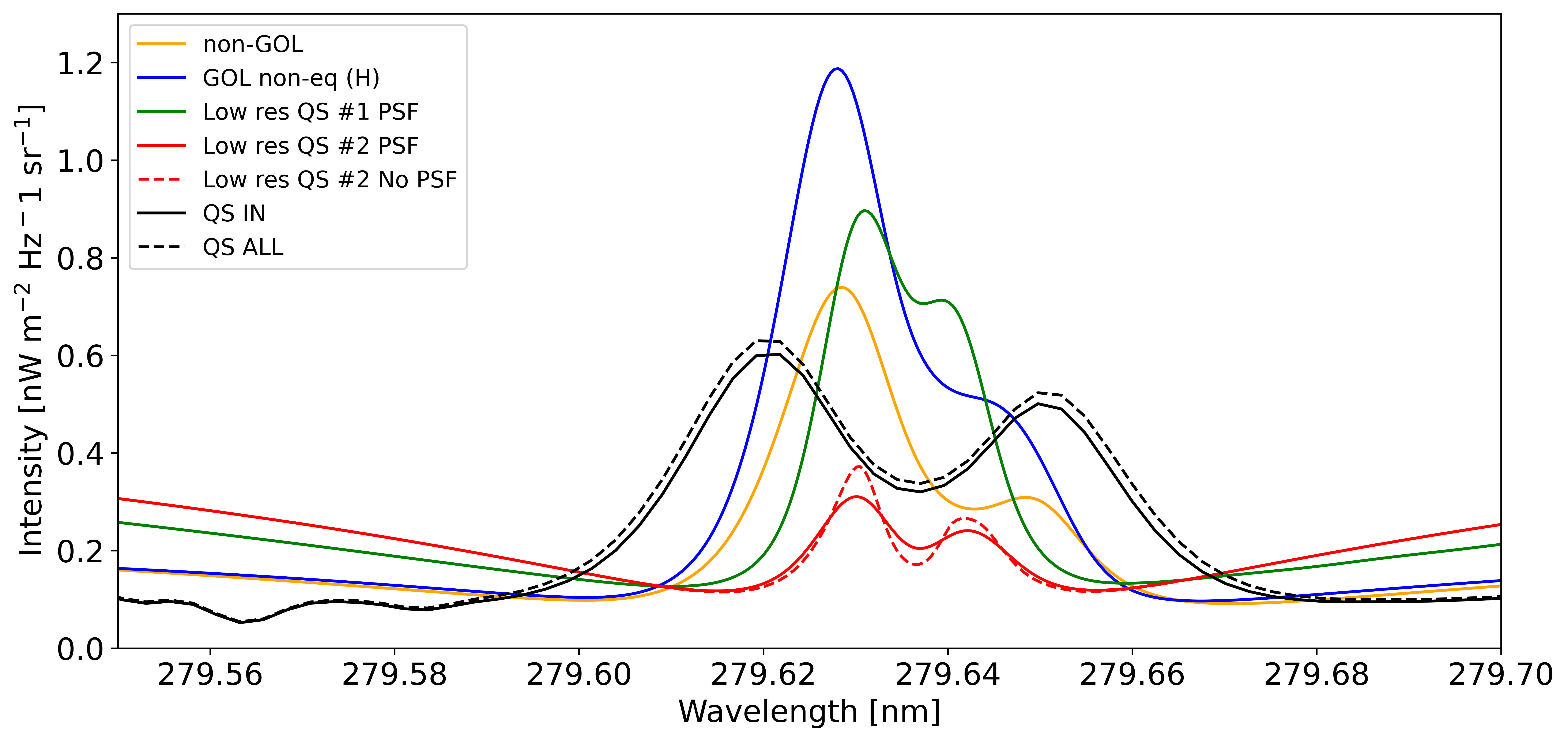}
	\caption{\label{fig:sim4km_profiles} \mgii\ profiles from a quiet Sun model made at high, 4~km, horizontal resolution, and including non-equilibrium hydrogen ionization and generalized Ohm's law. The dashed profiles represents IRIS observations taken in quiet Sun region where the solid line corresponds to IN \citep[see ][for details]{Martinez-Sykora2022_dkist}. For comparison a \mgii\ profile from the pre-emergence stage of the flux emergence model is added with 100~km resolution is shown in red.}
\end{figure}

Let us now compare these models with a model where the resolution is greater ($\sim 5$~km vs $\sim 100$~km) and with increased physical complexity: The {\tt QS nGOL} and {\tt QS GOL NEQ(H)} models. These models were initially  run for 51~minutes at which point the magnetic field statistically reached a steady state with $\uBz = 17$~Gauss, and $B_{\rm rms} = 57$~G, and with some flux concentrations reaching 2~kG at photospheric heights. These models have a simplified magnetic topology that mostly includes granular-scale magnetic fields (typical of internetwork regions) and do not include regions that resemble quiet Sun network.  In both models magnetic field is generated in-situ at chromospheric heights from conversion of kinetic energy to magnetic energy \citep{2019ApJ...878...40M}. In the model that includes the generalized Ohm's law and non-equilibrium hydrogen ionization, high rates of heating occur in regions where ions and neutral particles slip in relation to each other, i.e. the upper chromosphere, while non-equilibrium hydrogen ionization has the consequence that in colder regions the electron density remains much higher than when treating hydrogen ionization in LTE. Figure~\ref{fig:qs_tgstruct_gol} shows the chromospheric temperature structure of the {\tt QS GOL} and {\tt QS GOL NEQ(H)} models, including the locations of the height of $\tau_\nu$=1 layers for the \kthree\ minimum and \ktwo\ peaks.

In Figure~\ref{fig:sim4km_profiles} we show the average \mgii\ profiles obtained from these high resolution quiet Sun models and compare them with the quiet Sun models discussed above  and with \iris\ quiet Sun and network observations. The low resolution profile is from a snap shot later in time in {\tt nw072100} where the line asymmetries are reversed and correspond better to what is usually observed. We find that the high resolution models, ({\tt QS nGOL} and  {\tt QS GOL NEQ(H)}) have intensities somewhat higher but of the same order as those observed. However, the profiles are asymmetric, with \ktwov\ substantially brighter than \ktwor\ in both models, perhaps indicating rising coronal temperatures and densities with associated flows. The core profiles are broader than the previous low resolution, non-GOL, LTE ionisation {\tt nw072100} and {\tt qs072100} models, and though they are indeed a closer match to observed quiet Sun profiles we still find that the profiles are narrower than what is observed. The introduction of higher spatial resolution seems to have half the discrepancy (with the observations) in line width, with {\tt QS GOL NEQ(H)} giving marginally broader profile. The synthetic intensity in this model gives a radiation temperature of $T_{\rm rad}\approx 5000$~K and a line width of 0.040~nm FWHM, as compared to 0.051~nm for IN, and 0.056 for NW observations presented above.

\begin{figure}
    \includegraphics[width=0.49\textwidth]{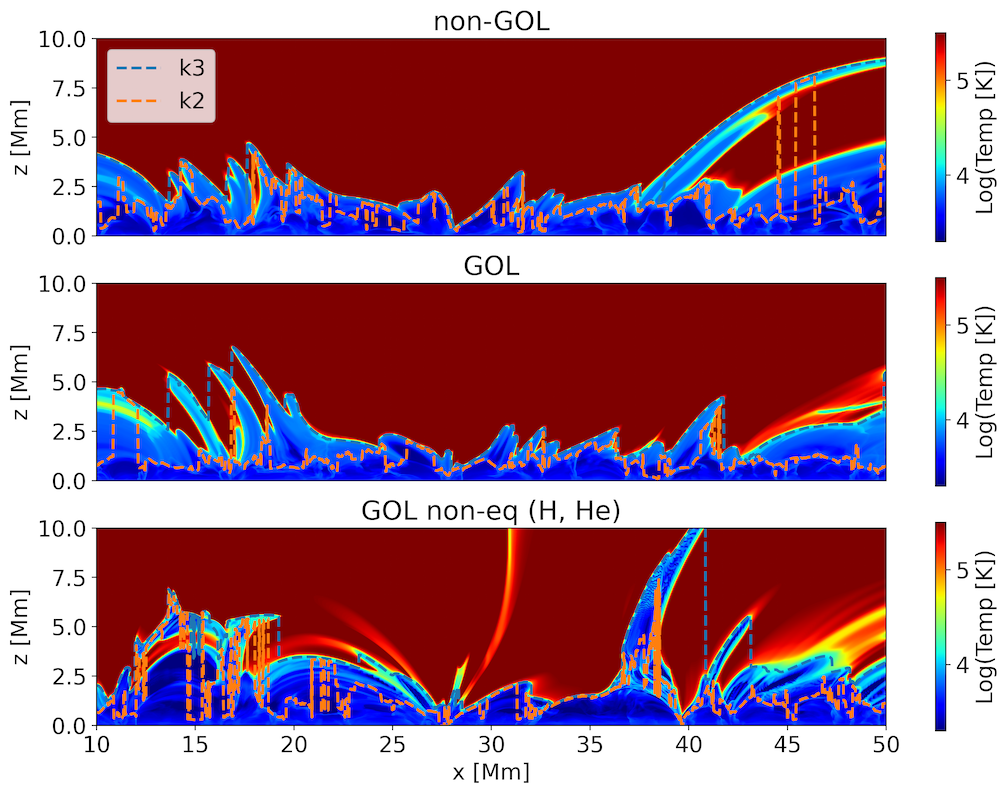}
    \caption{\label{fig:spicule_tgstruct_gol} Temperature structure of 2.5D spicule models run at high resolution (14~km) as standard MHD (top panel), with generalized Ohm's law (middle panel), and with generalized Ohm's law and non equilibrium hydrogen and helium ionization. The heights of the $\tau_\nu=1$ layers for the \ktwo\ and \kthree\ spectral locations are shown as dashed lines  }
\end{figure}

\begin{figure}
    \includegraphics[width=0.43\textwidth]{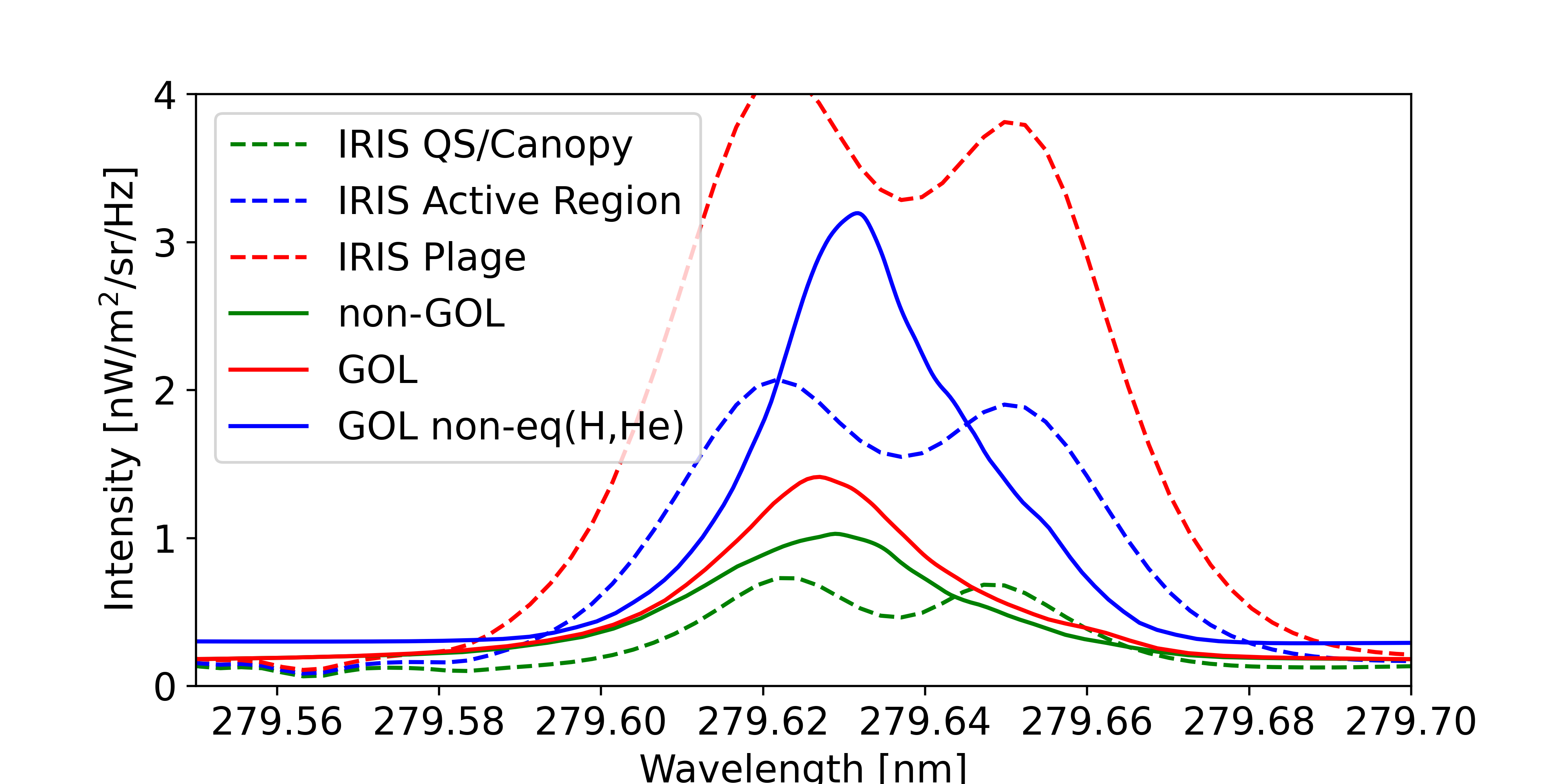}
	\caption{\label{fig:spicule_profiles_gol_hhe} Profiles from the ``spicule model'', which is 2D but made at high, 14~km, horizontal resolution. The red and green lines show the non-GOL and GOL models respectively, while the blue line shows the profile resulting from a model incorporating both GOL and non-equilibrium hydrogen and helium ionization. The dashed lines are profiles from IRIS observations of a quiet Sun and active region.}
\end{figure}


\begin{figure*}
\includegraphics[width=0.98\textwidth]{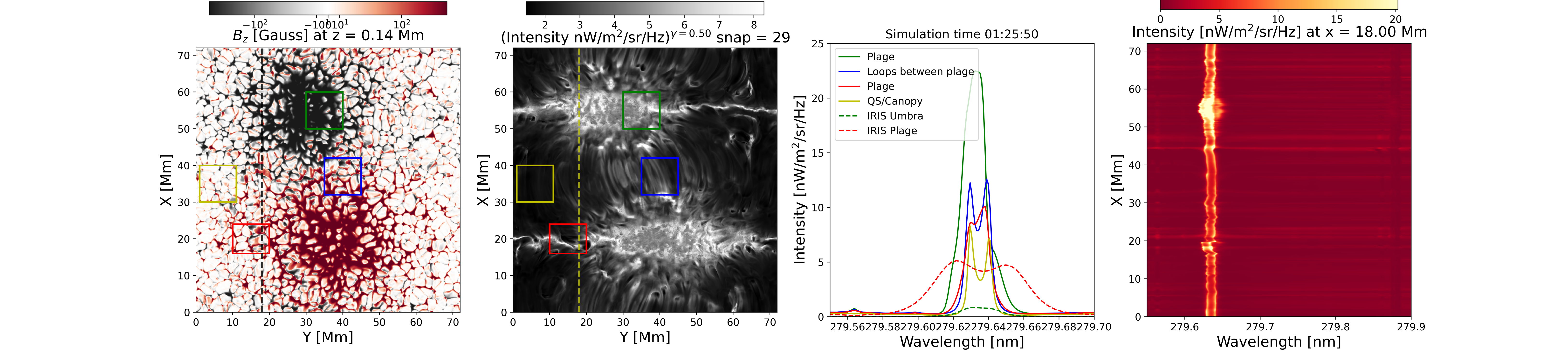}
\caption{\label{fig:plage_panels} The left panel shows a synthetic magnetogram of the simulation box, scaled to $\pm 750$ Gauss, while the remaining panels shows the \mgii\ k line core (second from left), average line profiles (second from right), and profile at $x=32$~Mm. The average profiles are computed in regions of interest delineated by red, green, blue, panels, and yellow, boxes. }
\end{figure*}

\subsection{Spicule model\label{sec:spicule}}

In the upper chromosphere, plasma $\beta << 1$ and the energetics and dynamics of the chromosphere become increasingly dominated by the magnetic field with increasing height above the photosphere. This is particularly true for regions of stronger field such as those characteristic of spicules. To investigate the effect of stronger, more dynamic fields, capable of exciting spicules, on the \mgii\ profile emission, the 2.5D atmospheres described by \citet{2017Sci...356.1269M,Martinez-Sykora:2020ApJ...889...95M} are utilized. 

This simulated atmosphere spans the upper convection zone $2.8$~Mm below the photosphere to the corona $40$~Mm above the photosphere.
The model also has a large horizontal extent, spanning $96$~Mm. A uniform spatial resolution of $dx=14$~km along the horizontal axis and a non-uniform resolution of up to $12$~km in the vertical direction ensures that the relevant physics is well resolved. Thus, though the model is 2.5D, convective motions in and below the photosphere are large enough to cause braiding of the magnetic field and a self-consistently heated chromosphere and corona through Joule heating, where we find coronal temperatures up to $2$~MK. At the resolution used, ambipolar diffusion is larger than the artificial numerical diffusion by 3 to 5 orders of magnitude in extended regions of the chromosphere. The magnetic configuration allows the formation of structures that closely resemble spicules of type~{\sc ii}, especially in the simulations that include GOL, as discussed in \citet{2017Sci...356.1269M}.  

The initial magnetic field in this model contains two plage-like regions of high magnetic field strength and opposite polarity. These are connected by magnetic loops that are up to ~50 Mm long, some footpoints of which are shown in Figure~\ref{fig:spicule_tgstruct_gol} which also shows the temperature structure of the chromosphere and transition region for all three models. The mean unsigned value of the vertical magnetic field $\uBz=86$~Gauss, $B_{\rm rms}=271$~Gauss, with magnetic flux concentrations reaching up to 1400~Gauss. 

Figure~\ref{fig:spicule_profiles_gol_hhe} shows the \mgii\ profiles resulting from this model when run with ``simple'' MHD, with GOL, and with GOL and non-equilibrium hydrogen and helium ionisation \citep[see also][]{2021A&A...654A..51B}. The profiles are computed in a $\sim3000$~km wide region containing the strongest magnetic fields, which is the site of strong spicule formation. The \ktwo, and \kthree\ intensities are stronger than that found in the quiet Sun, and though slightly weaker, are approaching the intensities found in active regions. This is especially true for the model where both GOL and non-equilibrium ionisation are considered. All three profiles are strongly asymmetrical, more so than the observed active region profiles, with the \ktwov\ peak significantly brighter than \ktwor. The synthetic line widths also approach or surpass those measured in the observed quiet Sun, with widths of $0.04$~nm FWHM in the non-GOL and GOL models, and $0.05$~nm FWHM in the GOL and non-equilibrium hydrogen and helium ionisation model. However these widths are still slightly narrower than that found in typical active regions and plage as shown in Figure~\ref{fig:iris_obs}.

\subsection{Plage model \label{sec:plage}}

When strengthening the magnetic field and expanding the areas covered by unipolar field, the dynamics of near surface convection, the structure of the chromosphere and corona, and the heating processes that determine the thermal and dynamic structure of the outer atmosphere are altered. A model typical of plage or small active region, named {\tt pl072100} in the following, illustrates these issues. This model is initiated with two opposite polarity patches of vertical magnetic field covering a large percentage of the modeled photosphere. The model has the same dimensions and is run with the same resolution as the large scale quiet Sun simulations, {\tt nw072100} and {\tt qs072100} discussed above. In contrast to the quiet Sun models, the field in {\tt pl072100} is quite strong, with an average unsigned vertical magnetic field strength of $\uBz=180$~Gauss. The field concentrations of the ``plage'' regions are of order $\pm 1800$~Gauss, but slightly weaker in the positive polarity plage. Though the field was originally confined, convective motions have caused the diffusion of quite strong fields also into the more quiet areas of the simulation at the time of the featured snap shot.  

The left panel of Figure~\ref{fig:plage_panels} shows the geometry of the photospheric magnetic field. The \mgii\ \kthree\ intensity is featured in the second panel of this figure and shows bright speckled areas in locations overlying the strong magnetic plage that appear quite similar to those observed (see Figure~\ref{fig:iris_obs}). The third panel contains average \mgii\ core intensities over four boxes covering regions of interest (ROIs) with different magnetic field configurations, placed both in regions of the strongest magnetic field as well as in areas between and away from the two main polarities. We find very bright emission in ROIs located in the strongest plage; with intensities $> 20$~\nanow\ which is much greater than the typical measured plage intensities of 5~\nanow.

The typical observed plage intensity translates to a radiation brightness temperature of $T_{\rm rad}\approx 6300$~K, while ranging from average radiation temperatures of $T_{\rm rad}=6100$~K for the \kthree\ core in the ``QS/Canopy'' patch through $T_{\rm rad}=7000$~K for the (blue box) region between the two polarities, to $T_{\rm rad}=7700$~K for the brightest, plage like, patch. The latter profile is single peaked and the small scale structure of the emission is similar to what is observed in observed plage. However, when we consider the line widths, we again find that they are all much narrower than what is observed, of order $0.025$~nm FWHM, which is less than half what is observed. Instead, the synthetic widths are very close to what is found in sunspot umbrae as can be seen by comparison with Figure~\ref{fig:iris_obs}.
Comparisons should also be made with the profiles found in the of the models seen in Section~\ref{sec:spicule} which has strong fields, but at higher resolution and with other physical processes at work. Likewise, we find strong field regions and single peaked profiles in models that include flux emergence as described in the following section.

\subsection{Emerging field models\label{sec:emerging_flux}}

The two low resolution quiet Sun models presented above had line core widths significantly narrower than what is observed, but also average unsigned magnetic field strengths of only $\uBz\approx 30$~Gauss. The line widths were greater in the quiet Sun model that was run at high resolution and with GOL and non-equilibrium hydrogen ionisation and a stronger, local dynamo generated, field, but even at this resolution ($5$~km) did not reproduce the observed widths. Let us now reconsider the low resolution case, but in which the field strength gradually rises, e.g. in an emerging flux model.

\begin{figure}
    \includegraphics[width=0.49\textwidth]{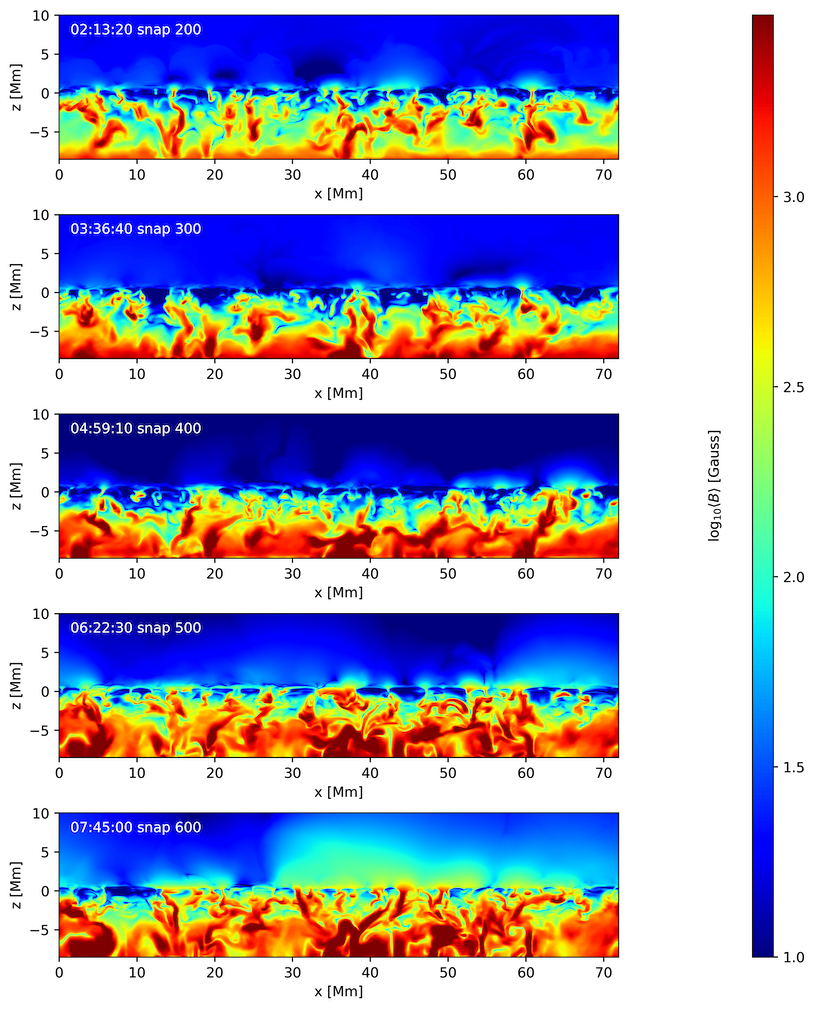}
	\caption{\label{fig:cz_fe} Evolution of the total unsigned magnetic field strength in the convection zone.}
\end{figure}

The emerging flux model is a continuation of the {\tt nw072100} quiet Sun model described above. Flux emergence is initiated by injecting into the original configuration a flux sheet, aligned with the $y$-axis, of strength  $B_y=200$~Gauss at the entire bottom boundary for 95 minutes. After this initial period the flux sheet strength was increased to, first $B_y=1000$~Gauss for 70~minutes and thereafter 2000~Gauss for another 150~minutes. Afterward, the strength of the injected field was reduced to $B_y = 300$~Gauss, which is injected continuously thereafter. Figure~\ref{fig:cz_fe} shows the evolution of the field in the middle of the computational domain ($y=35$~Mm) over an 8~hour period. The initial 100~Gauss field is kneaded and pulled down by convection below the photosphere in several locations, while some field rises through the chromosphere and into the corona in the first few hours of the simulation. In the corona the field is nearly horizontal and the coronal plasma becomes quite hot, only cooling slowly as the field diffuses through radiative cooling and losses through conduction. The latter is hampered by the weak thermal connection between the corona and chromosphere.

This early coronal field has a strength of some 20~Gauss, decreasing from the fourth to the fifth hour to a minimum of 10~Gauss. From the fifth hour onward, portions of the strong injected flux sheet reach the photosphere, rising to the upper atmosphere layers above, in places where the emerging field is strong enough in the photosphere to overcome its lack of buoyancy \citep{1979SoPh...62...23A,2004A&A...426.1047A}. 

\begin{figure}
    \includegraphics[width=0.49\textwidth]{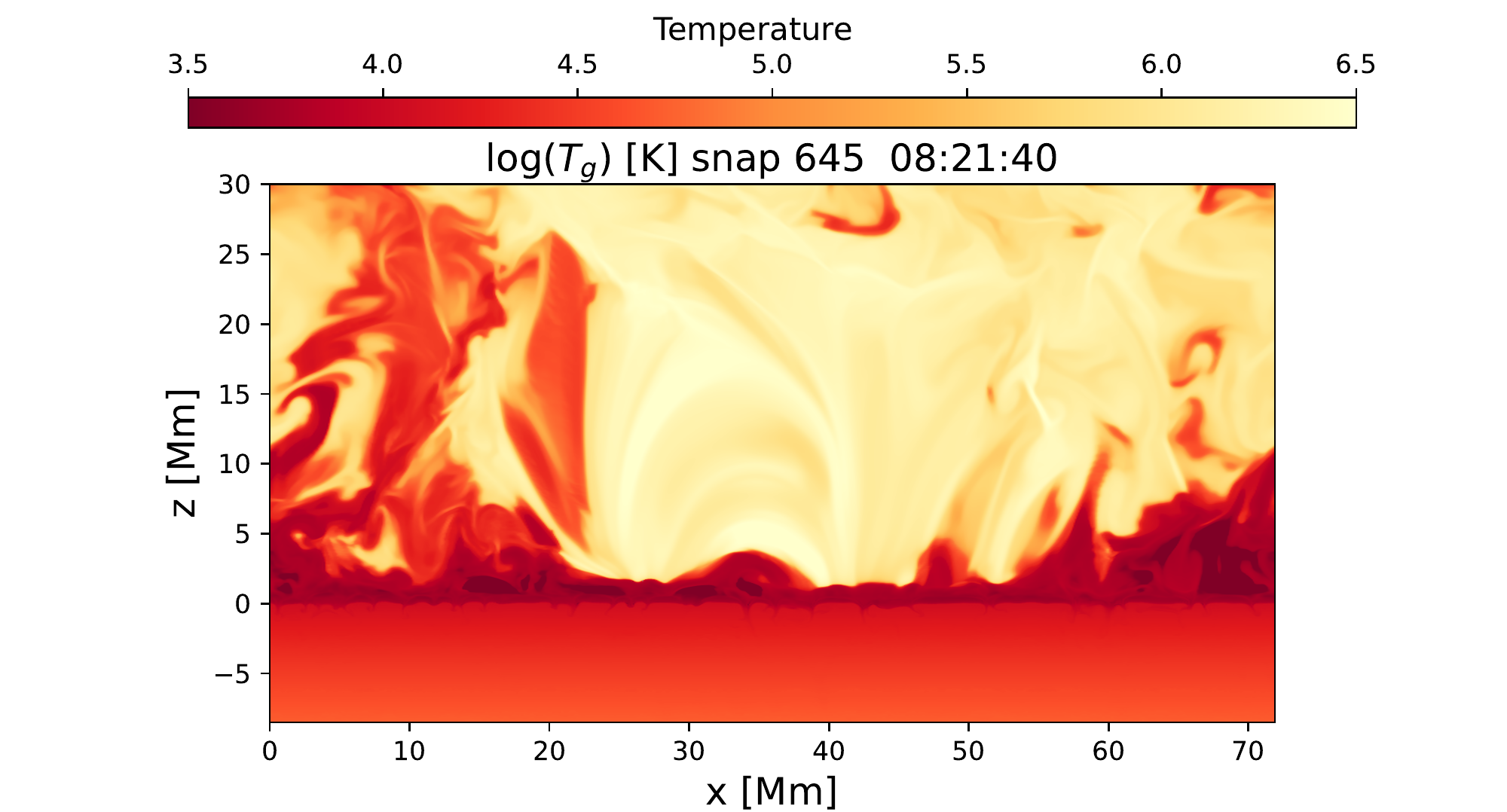}
	\caption{\label{fig:cbp_tgstruct} Temperature structure at $y=32$~Mm at time $t=$~\htime{8}{21}  in the ``flux emergence'' ({\tt nw072100}) at a stage where flux emergence is more or less complete.}
\end{figure}

The expanding and increasing coronal field leads to both a restructuring of the coronal field topology, the introduction of cool plasma carried along with the field, and rising coronal heating rates. The temperature structure in the model towards the end of the run (at 8~h~21~m) is shown in Figure~\ref{fig:cbp_tgstruct}. As in the quiet Sun models, we see that chromospheric temperatures are largely confined to the 2~Mm above the photosphere. However, there are several regions where cool, $T_g<30\,000$~K gas is present up to 5 or even 10~Mm above the photosphere. This cool gas having been carried up into the corona by the magnetic field as it expands from the photosphere, and now forming cool fibrils. Furthermore, in regions where heating is strong, we find loop shaped structures with temperatures $T_g>5$~MK where the chromosphere is compressed and the transition region temperature rise occurs already at 1.5~Mm above the photosphere, very similar to the structure we find in the plage model {\tt pl072100}.

\begin{figure*}
    \includegraphics[width=0.98\textwidth]{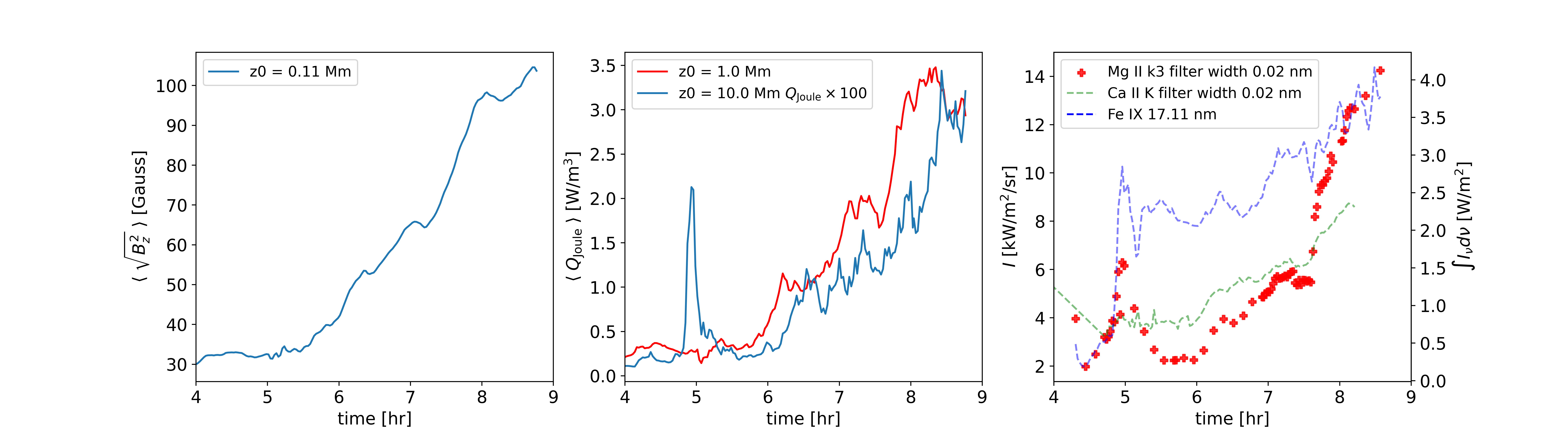}
	\caption{\label{fig:average_atm} Evolution of the average unsigned vertical magnetic field $\langle |B_z|\rangle$ (left panel), Joule heating in the chromosphere and corona ($\times 100$) at $z=1$~Mm and $z = 10$~Mm (middle panel), and average intensities of the \mgii\ k line core, the \ion{Ca}{2}~K line core and the \feix~17.1~nm line}
\end{figure*}

In Figure~\ref{fig:average_atm} the average field, coronal Joule heating and \ion{Ca}{2}~K, \ion{Mg}{2}~k, and \ion{Fe}{9}~17.1~nm line intensities are shown as a function of time, starting roughly an hour before the emerging field reaches and pierces the photosphere and for the next 5 hours. Five hours into the simulation run, emerging flux that has collected just below the photosphere breaks through and interacts with the ambient coronal field; this causes a spike in (coronal) Joule heating, which increases by more than a factor of three for a short interval, to $2\times 10^{-2}$~W~m$^{-3}$, with an accompanying spike in both the \mgii\ and \feix\ intensities. There is no corresponding spike in the \ion{Ca}{2} intensity, and we note that the chromospheric Joule heating does not increase at this time either, but rather remains at $Q_J=0.3$~W/m$^3$. 
After the spike the \feix\ emission remains high, while the \mgii\ intensities fall back to their previous value as soon as the Joule heating subsides. The mean vertical magnetic field strength $\uBz$ in the photosphere remains at some $30$~Gauss until 5.5~hours, at which point it rises with the increasing emergence of magnetic flux, passing 60~Gauss at 6.75~hours, and reaching 100~Gauss at the 8~hour mark at which point the mean strength remains nearly constant. 

Figure~\ref{fig:average_atm} attests that as the average magnetic field strength rises, so does the Joule heating rate, both in the chromosphere at $z=1.0$~Mm and in the corona at $z=10.0$~Mm (and in points in between). The chromospheric heating remains at $Q_J=0.3$~W/m$^3$ until hour 6, thereafter rising rapidly to $>3.0$~W/m$^3$ in the span of two hours. Except for the spike in heating around hour 5, mentioned above, the coronal heating rate mirrors the chromospheric rate, but at a level that is a factor 100 lower in amplitude. 

The increased heating rate impacts both the intensities of \ion{Ca}{2}~K and \mgii\ k  in the chromosphere and the \feix~171.1~nm line in the corona. We compute the average line core intensities of \ion{Ca}{2} and \mgii\ integrated over a 0.02~nm FWHM wide Gaussian filter and the total line intensity of the \feix\ line. The increase in the \mgii\ line intensity becomes especially apparent when the average field becomes larger than $\uBz\approx70$~Gauss, when we see a rapid rise in the core intensity. 

\begin{figure*}
    \includegraphics[width=0.98\textwidth]{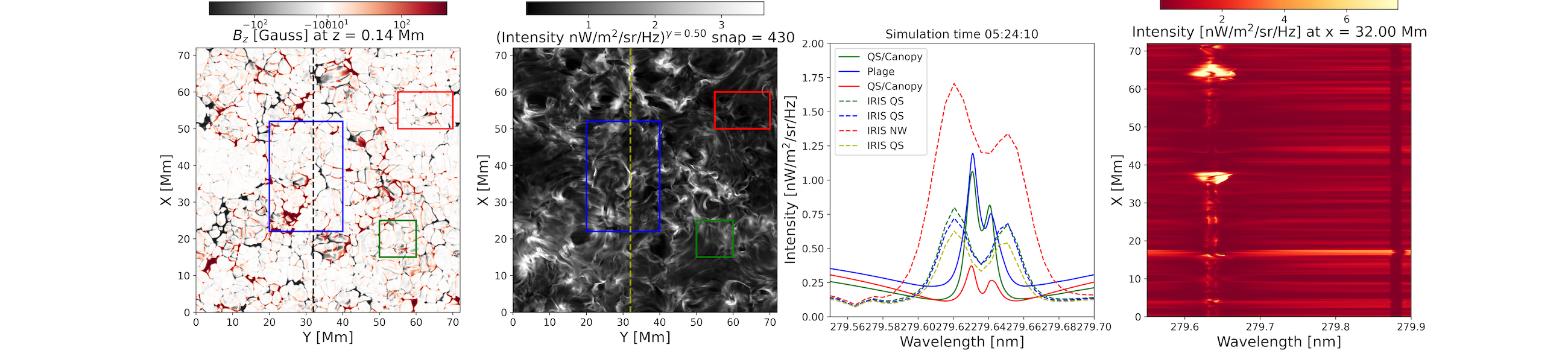}
    \includegraphics[width=0.98\textwidth]{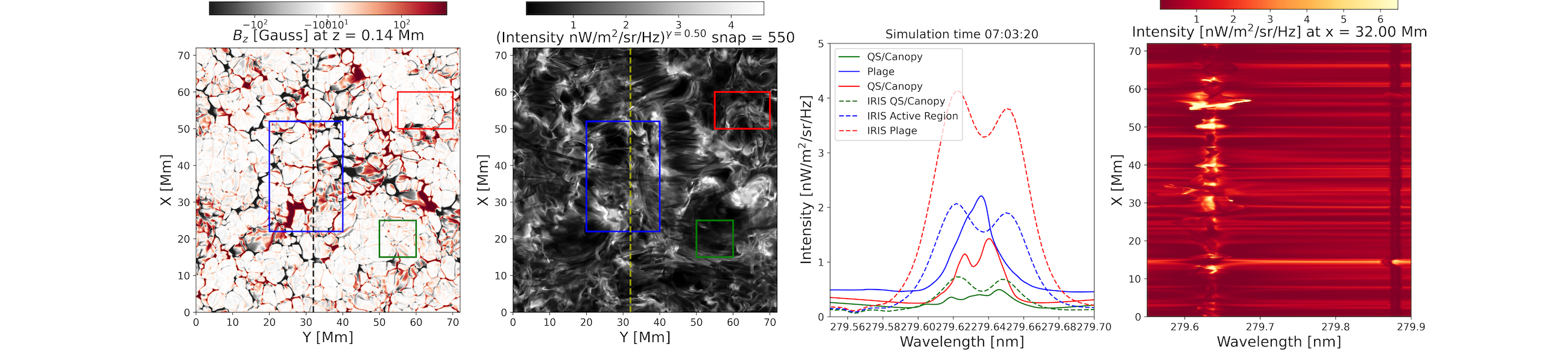}
    \includegraphics[width=0.98\textwidth]{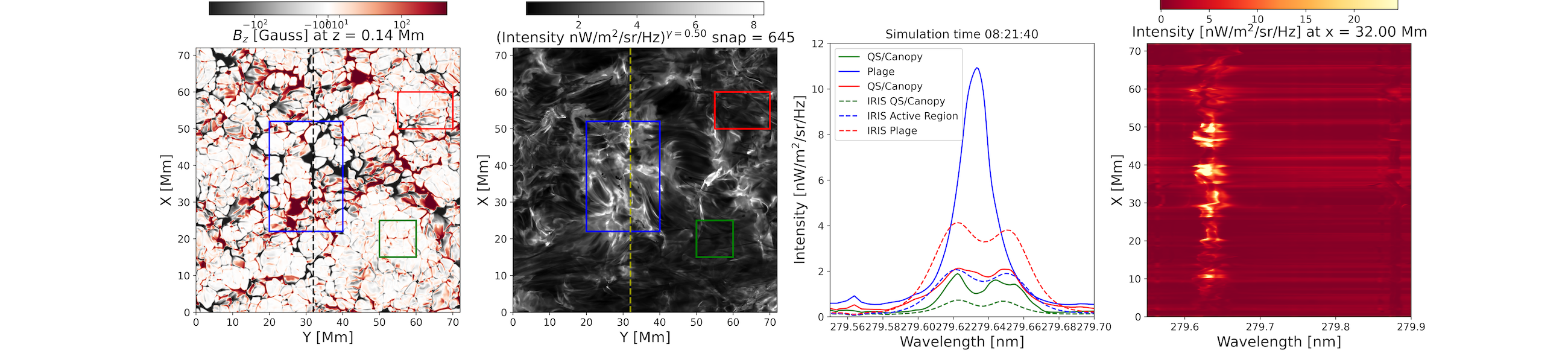}
	\caption{\label{fig:fe_profiles} ``Emerging Flux'' model profiles. The layout is the same as in Figure~\ref{fig:iris_obs} while, for comparison with the model, we have added the plage, active region, and quiet Sun profiles from the middle row of that figure as dashed lines.}
\end{figure*}

Let us now consider the \mgii\ profile in greater detail. In Figure~\ref{fig:fe_profiles} we find line-plots of \mgii~k as well as a map of the \mgii~k and the \mgii~2799 triplet line along $x=32$~Mm (dashed yellow line in the second panel from the left). These profiles are taken from periods just after the field has pierced the photosphere, and at later stages as field has risen into the outer atmosphere and pervades the steadily more magnetically active chromosphere and corona. 

The top row shows the \mgii\ line at $t=$~\htime{5}{23}, some 67~minutes after the previous {\tt nw072100} model ``quiet Sun'' snap shot shown in Figure~\ref{fig:qs_profiles}. The intensities at this time, one hour later, are roughly unchanged and of the same order as those obtained with the IRIS observations of quiet Sun regions. The \ktwo\ peaks are slightly asymmetric, but at this time with the \ktwov\ peak brighter than the \ktwor\ peak, as is often observed, and as opposed to the snap shot shown in Figure~\ref{fig:qs_profiles}. However, the line profiles are still too narrow to resemble those observed. We note that the vertical unsigned magnetic field at this point in time is still only slightly larger than 30~Gauss. 

The middle row shows the \mgii\ spectra at $t=$~\htime{7}{3} when the unsigned magnetic field strength has risen to $\uBz \approx 60$~Gauss. At this time we find that the \mgii\ core intensities (in two of the three ROIs shown) have increased by roughly a factor 2 --- and equivalent to the radiation temperature $T_{\rm rad}$ increasing by 400~K from 5250~K to 5650~K. The line widths have all increased significantly and are approaching the same widths as observed quiet Sun line widths. The profiles in the ``quiet'' regions outlined by red and green boxes are twin peaked (or complex), while the profile formed in the region of strongest field is single peaked and with a high, and rising, intensity of 2~\nanow. We note that given the amount of flux emergence in the simulation, it is not fully clear which type of solar region (in terms of observations) is the best comparison point. 

\begin{figure}
    \includegraphics[width=0.43\textwidth]{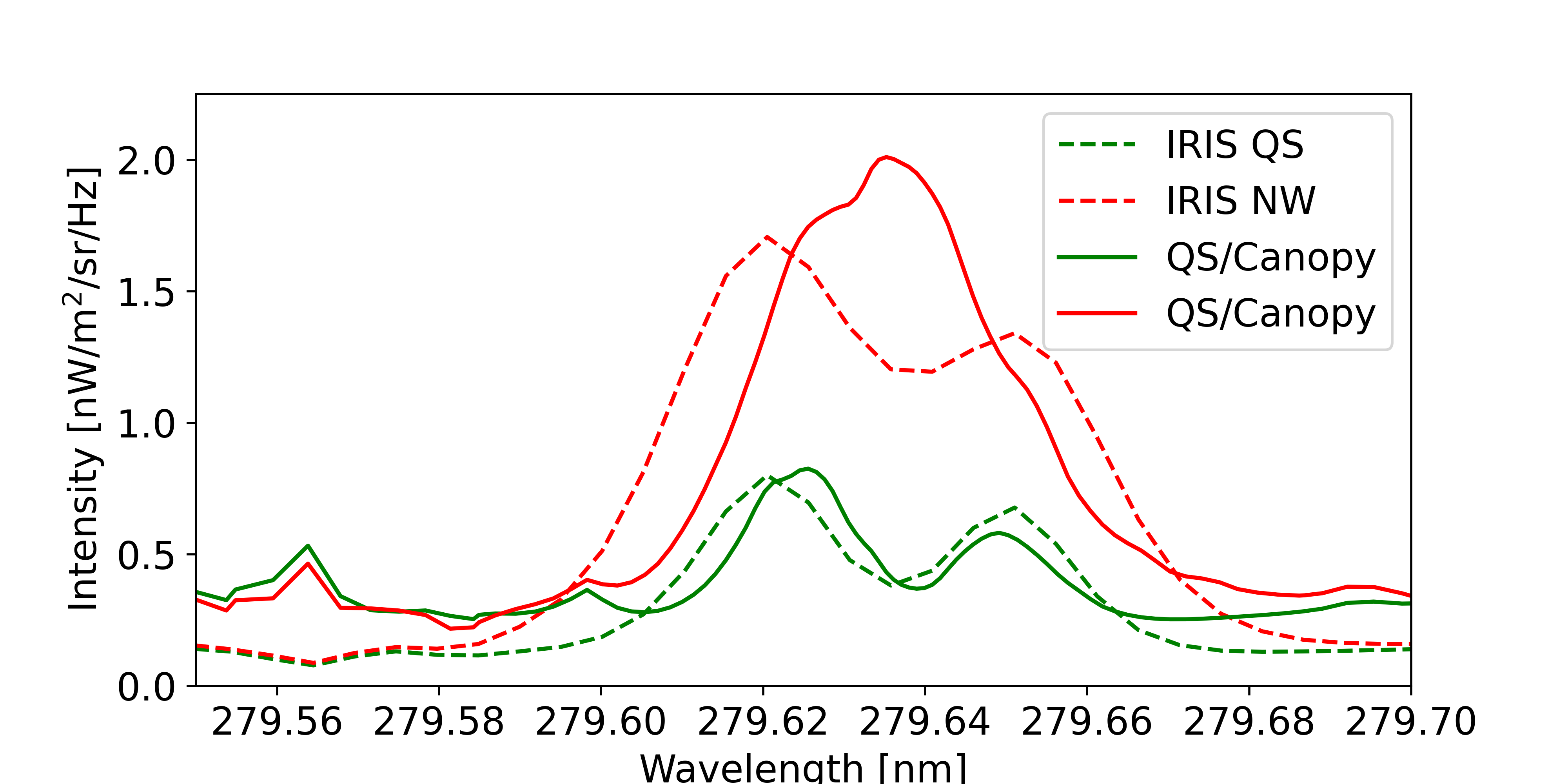}
    \includegraphics[width=0.43\textwidth]{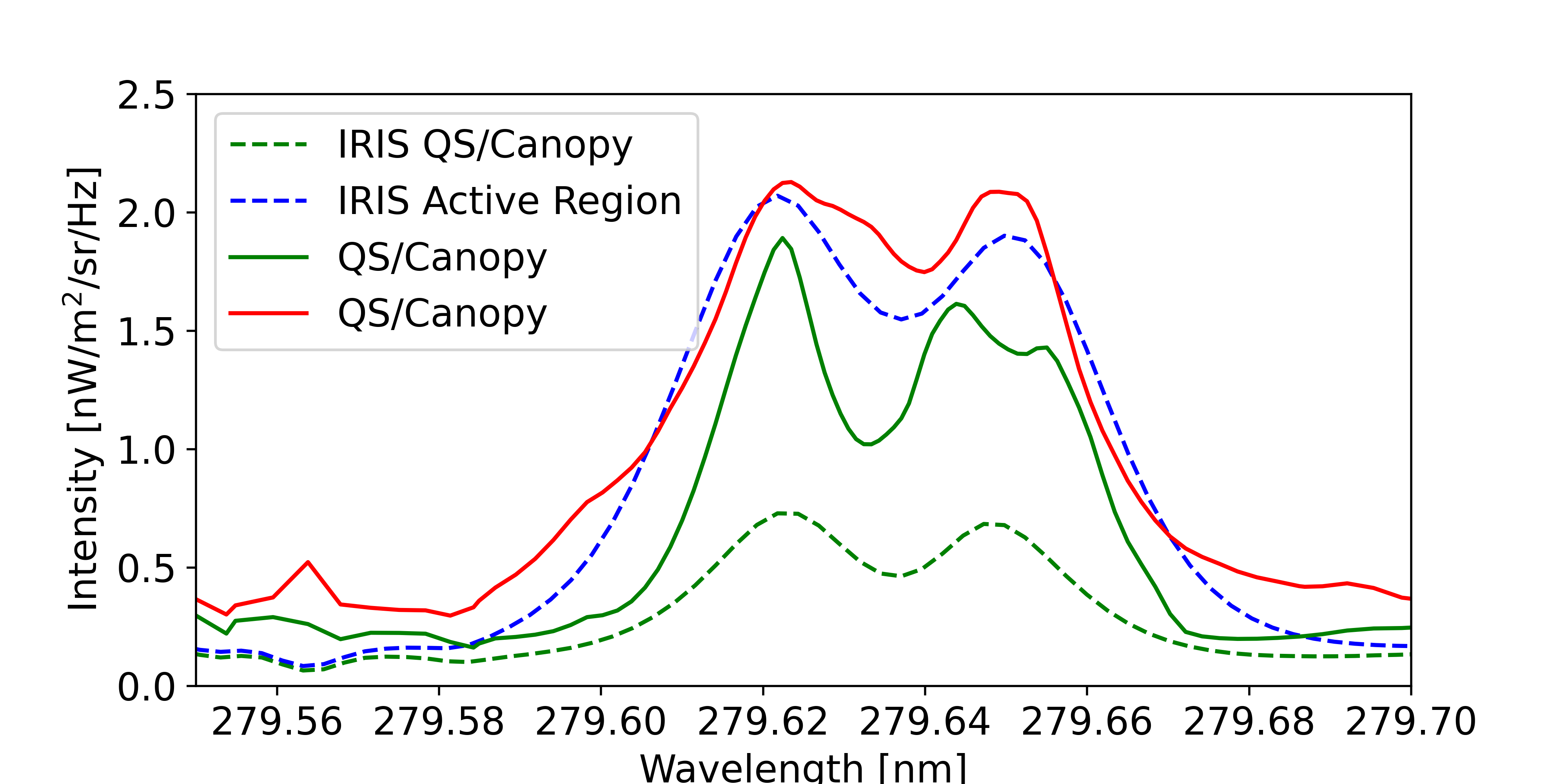}
	\caption{\label{fig:postfe_profiles} \mgii\ profiles from late stages, at \htime{7}{44}\ upper panel and \htime{8}{21}, of the flux emergence simulation at , in magnetically weaker regions. The profiles drawn in red and green come from the equivalently colored regions of interest shown in Figure~\ref{fig:postfe_profiles}.}
\end{figure}

Finally, in the bottom row of Figure~\ref{fig:fe_profiles} the profiles at simulation time $t=$~\htime{8}{21} are presented. At this time the magnetic field has reached $\uBz\approx 100$~Gauss with an accompanying increase in both chromospheric and coronal heating $Q_{\mathrm Joule}$, as is visible in the middle panel of Figure~\ref{fig:average_atm}. The two darker ROIs, outlined with red and green boxes, are centered on darker ``quiet Sun'', or canopy, areas which both have line cores which are as wide as is observed with intensities that lie between those found in observed average quiet Sun and active regions; i.e. between 0.5 and 2~\nanow. The ROI which covers the region of most intense activity, outlined by the box drawn in blue box shows a very intense, of order 10~\nanow, single peaked average profile that is wide, though not as wide as the ``quiet Sun''/canopy profile derived from the area covered by the red box ROI. These darker profiles, along with the equivalent profiles from $t=$~\htime{7}{45} are shown in greater detail in Figure~\ref{fig:postfe_profiles}.

\section{Formation of \mgii\ \label{sec:magi_formation}}

\begin{figure}
\includegraphics[width=0.49\textwidth]{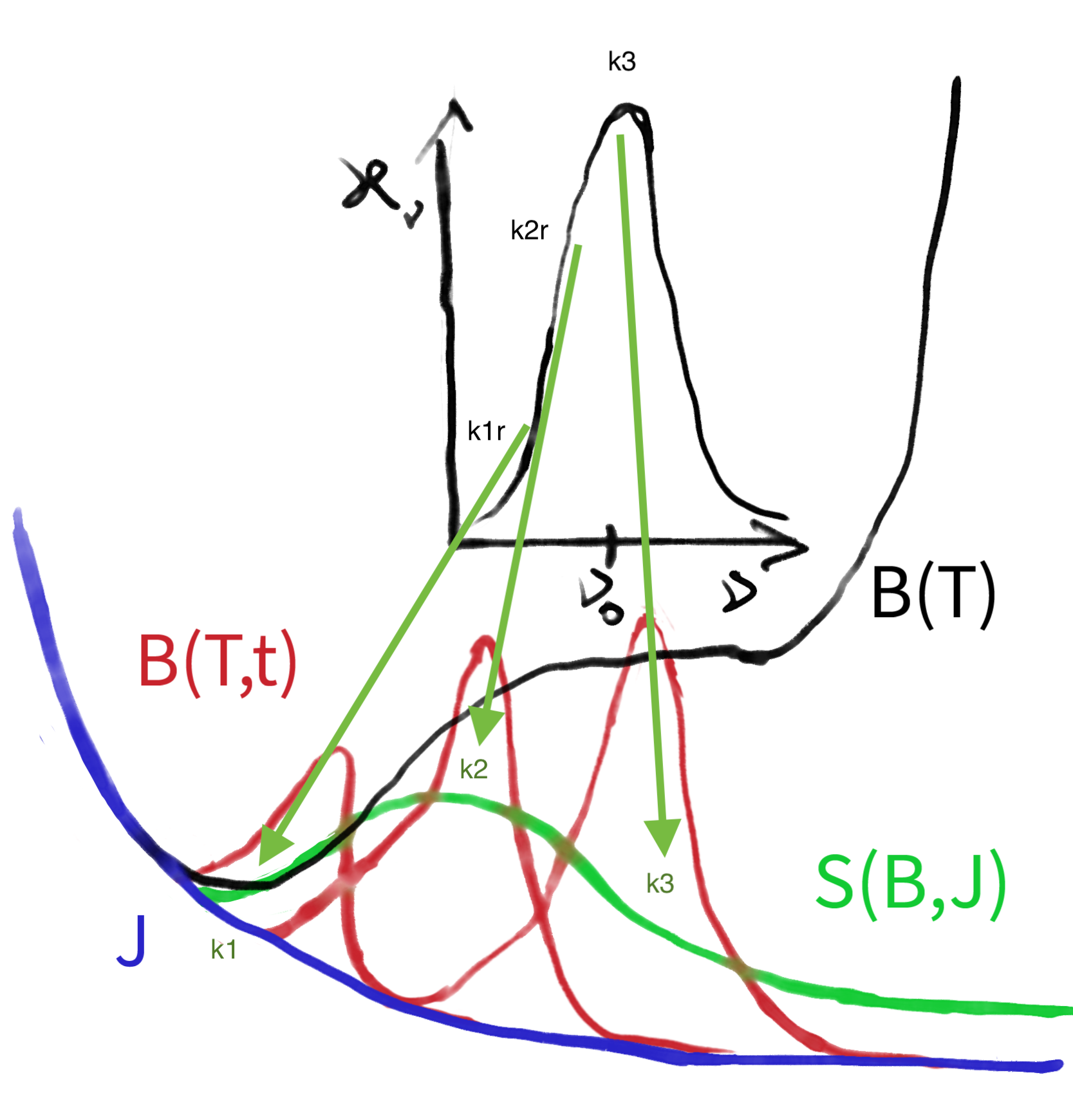}
\caption{\label{fig:source_function} Schematic cartoon of the formation of \mgii. The Plank function $B$ ($\equiv B_\nu(T_g)$ in the main text) as function of height is shown in black; the red lines indicating the possible non-stationarity of the atmosphere in cases where large amplitude compressive waves are important. The mean intensity $J$ is plotted in blue, while the source function $S(B,J)$ is plotted in green. The opacity $\kappa_\nu$ as function of frequency is plotted in black. The \mgii\ line profile will reflect the source function such that $I_\nu\approx S(\tau_\nu=1)$ and the locations of $\tau_\nu = 1$ are indicated for the \kone, \ktwo, and \kthree\ spectral features.}
\end{figure}

\begin{figure*}
\includegraphics[width=0.98\textwidth]{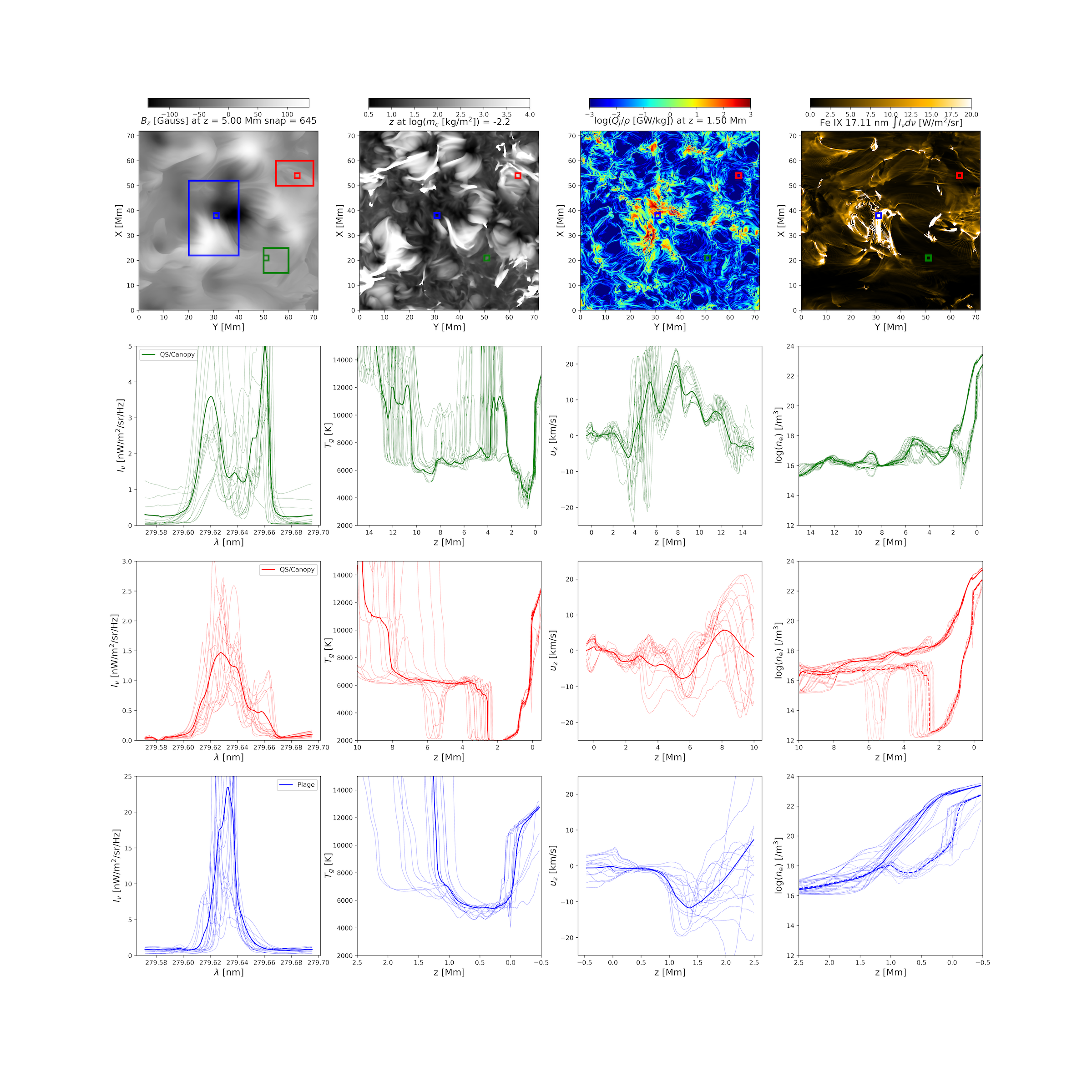}
\caption{\label{fig:mg_prof_atmos} Top row: The vertical magnetic field $B_z$ at 5~Mm above the photosphere, height of column mass $\log(m_c])=-2.2$ (a \kthree\ proxy), Joule heating per unit mass at 1.5~Mm above the photosphere and the total intensity of the \feix\ spectral line at time \htime{8}{21}\ in simulation {\tt nw072100}. The three following rows show average and individual \mgii\ line profiles, the temperature, vertical velocity, and hydrogen and electron particle densities as functions of height in restricted regions placed within the larger regions of interest used in earlier figures. }
\end{figure*}

The formation of \mgii\ and in particular the details of how the intensity and width of the line core, 
\kone, \ktwo, and \kthree, are formed is quite complex and dependent on several aspects of the chromosphere, corona, and magnetic field, both along the radiating ray and in the general vicinity of the emitting plasma. These include the chromospheric density and temperature structure, velocity flows and turbulence, which all can and will alter the opacity and emissivity of \mgii. The formation of \mgii\ is discussed in detail in \citet{2013ApJ...772...90L}. Furthermore, the root cause of ``opacity broadening'' is discussed in depth in \citet{2015ApJ...811...80R}. We here repeat some of the general formation properties most relevant for the context of this paper.

The emergent intensity $I_\nu$ of a spectral line can be approximated by the Eddington-Barbier relation

\begin{equation}
I_\nu(\mu=1)=\int_0^\infty S e^{-\tau_\nu}\kappa_\nu dz \ \approx S(\tau_\nu=1)
\end{equation}

\noindent where $S$ is the source function, $\kappa_\nu$ the opacity, $dz$ a length element along the ray, and $\tau_\nu$ is the optial depth at frequency $\nu$. In principle, the source function is a function of frequency as well for lines in which partial frequency distribution is important, such as outside the \ktwo\ peaks of the \mgii\ line, but for the purposes and context of this discussion, we assume this can be ignored. Also ignored are the effects of horizontal radiative transfer, which is important for the core of the \mgii\ h\&k lines \citep{2013ApJ...772...90L}. 

Instead, let us concentrate on the frequency independent source function as a function of height in the outer solar atmosphere. Deep in the atmosphere, the mean free path for (all) photons in the line is very short, essentially no photons escape, and conditions are very close to local thermodynamic equilibrium, such that
\begin{equation}
S = J = B_\nu(T_g)
\end{equation}
where $J$ is the mean intensity and $B_\nu(T_g)$ is the Planck function, which depends on the local gas temperature $T_g$. In general the source function will depend on both the local temperature through $B_\nu$ and the radiation field $J$. As one moves outward in the atmosphere, $B_\nu$
will follow the chromospheric temperature structure while the mean intensity $J$ will decrease as photons can more readily escape the atmosphere. The source function will initially follow $B_\nu$, but will eventually fall towards $J$ as the radiation field and local emissivity decouple. 

Magnesium is an abundant element with \mgii\ being the dominant ionization state in the chromosphere at temperatures below some
15~kK. The line core (\kthree) is therefore formed close to the transition region temperature rise \citep[see Figures~\ref{fig:qs_tgstruct_gol} and \ref{fig:spicule_tgstruct_gol} or][]{2013ApJ...772...90L}. At that point, the source function is often sufficiently decoupled from the Planck function such that the value of the source function is higher further down in the atmosphere. The \ktwo\ peaks are formed at a wavelength separation from the line core such that $\tau_\nu=1$ falls at this local
maximum of the source function. 
For a larger column mass difference between the locations of the $\tau_{k_3} = 1$ and $\tau_{k_2} =1$ points, we need to go
further out in the absorption profile to have an opacity low enough and we get a broader intensity profile.

A high density at the height where the \kthree\ is formed will ensure that $S$ is close to $B_\nu(T_g)$ and a high temperature in the upper chromosphere will give a high intensity. This will, for example, occur if the coronal temperature is high, $> 2-3$~MK, and thermal conduction forces the transition region to small geometric heights and greater densities as is the case in the plage model shown in Figure~\ref{fig:plage_panels} and the later stages of the flux emergence model where a small coronal bright point with associated high coronal temperatures is forming. On the other hand, a coronal temperature $<1$~MK leads to a transition region located at large heights, $z>3$~Mm above the photosphere, and a source function $S$ that is decoupled from the Planck function and approaching the steadily decreasing mean intensity $J$,  leading to low \kthree\  intensities.  


Note that it is not only the coronal temperature that can modify the transition region geometry and density; a strong horizontal magnetic field with associated Lorentz force will do the same, for example in the form of spicules, or low-lying loops, some forming the fibrils seen in H$\alpha$, the \ion{Ca}{2}~K\&H lines, or \mgii~h\&k. Likewise, emerging flux can raise cold photospheric material to great heights, modifying the chromospheric geometry, while at the same time being the source of enhanced reconnection activity and heating as the emerging field comes into contact with the pre-existing ambient field. 

The \ktwo\ intensity is usually higher than \kthree\ as it is formed at the height where $S$ is still strongly coupled to $B_\nu(T_g)$ and thus the chromospheric temperature rise. There are two ways to obtain a single peak profile. If the upper chromosphere density is so high that the source function still remains coupled to the local temperature. there will be no local maximum of the source function
and therefore no \ktwo\ peaks and a ``single peaked'' profile.
An alternative could be for cases (e.g., in the umbra, where faint single peaked profiles are common) where the density is low throughout the chromosphere so that the source function is never well coupled to the local temperature, so that there is no local peak in the source function with frequency.

The \kone\ spectral feature, the point of minimum intensity just outside the core, is found at the frequency where photons can escape readily from the chromospheric temperature minimum, some few hundred kilometers above the photosphere. A schematic cartoon of the relationship between $S$, $J$, $B_\nu(T_g)$, and the \kone, \ktwo, and \kthree\ frequencies is presented in Figure~\ref{fig:source_function}. The Planck function $B(T_g,t)$ in that figure is added to illustrate that the temperature structure of the chromosphere can be highly time variable, and indeed, for the lower chromosphere the time variability can be its most salient feature \citep[e.g.,][and references cited therein]{2019ARAA..57..189C}. 

Two processes set the width of the \mgii~k core; broadening of the atomic absorption/emission profile due to small scale turbulent velocities in the chromosphere at the locations of core emission, and ``opacity broadening'' as also discussed by \citet{2015ApJ...809L..30C} and \citet{2015ApJ...811...80R}. The former could be the result of high frequency waves, motions driven by episodic heating events due to magnetic reconnection, or motions resulting from instabilities such as the Kelvin-Helmholtz instability \citep{2015ApJ...809...72A} or the Thermal Farley Bunemann instability \citep{2022arXiv221103644E}. However, the measured line width of the optically thin \ion{O}{1}~135.4~nm line in plage regions  \citep{2015ApJ...809L..30C,2015ApJ...813...34L} is only of order 10--15~km/s FWHM\footnote{It is important to note that for the interpretation of line broadening in terms of motions in the solar atmosphere, a comparison with the 1/e width should be performed since that provides the most probable velocity, from a statistical point of view. To convert our FWHM values into 1/e width, a division of FWHM by 1.67 should be applied.}, which is not enough to explain the \mgii~k core width. We note though that implicit in this argument is the assumption that \ion{O}{1} is formed in the same region of the atmosphere as \ion{Mg}{2}. While numerical modeling of quiet Sun suggests that this is true \citep{2015ApJ...813...34L}, it is not clear whether that is the case for all regions on the Sun. Another way of obtaining a broad spectral core profile is by increasing the vertical extent of the dense, chromospheric temperature plateau. This will form a broad range of frequencies in which the source function is coupled to the local temperature and only far from the line center, $\nu_0$, will $S$ decrease to half of
the peak value (where the FWHM intensity is formed).

Thus, in order to reproduce the observed intensities and line core widths of the \mgii~k line, a solar chromosphere that is hot and dense over an extended range in height, or more accurately in optical depth, is required. In quiet Sun models, we do find an increase in core width in higher resolution models, as shown in Figure~\ref{fig:qs_profiles}, which could be caused by increased turbulent velocities and/or increased mass loading of the upper chromosphere because of more concentrated energy release and resulting stronger motions. It is also possible that the topological differences between the various QS  models play a role in the different line widths. A further increase of the width is found when considering models where GOL, and especially GOL and non-equilibrium hydrogen (and helium) ionization are included. The GOL models feature higher average chromospheric temperatures, larger chromospheric scale heights and hence greater mid and upper chromospheric densities. However, while models including GOL and NEQ ionisation do increase the width of the line core reducing the discrepancy significantly, neither of these models sufficiently change chromospheric structure enough to match observed quiet Sun \mgii\ core widths.

Models of the more active Sun, in which the magnetic field plays a more prominent role, are also capable of producing larger line core widths (Figure~\ref{fig:spicule_profiles_gol_hhe}), especially when GOL and non-equilibrium hydrogen and helium ionization are included. Spicule dynamics bring significant mass up into the upper chromosphere in these simulations. This increases the density, while currents associated with the spicule acceleration heat the upper chromosphere, transition region and lower corona, leading to the atmospheric structure needed to produce large line widths.  Yet, while the synthetic line core widths from these models are of the same order as what is observed in the quiet Sun, they are not quite wide enough to reproduce the line profiles seen in active regions or plage. We note though that these models are 2D models, and it remains unclear whether expansion into 3D would further reduce the discrepancies.

\begin{table*}[tbp]
\centering%
\begin{tabular}{ cccccc } 
IRIS observation/ & Region type & Intensity & Width & Resolution & Comment \\
Model name & & max(I) [nW/m$^2$/Hz/sr] & FWHM [nm] & [km] & \\
 \hline
Sun Center  & QS & $1.0$ (IN) & $0.051$ (IN)  & & \ktwov\ stronger \\
&&1.75 (NW)& $0.056$ (NW) & &  Figure \ref{fig:iris_obs} \\
NOAA 12296 & Plage/ & $>4$ (Plage) & $0.059$ & & \ktwov\ stronger \\
&small AR&$2.0$ (AR)&$0.059$& & Figure~\ref{fig:iris_obs} \\ 
NOAA 12480  & AR/ & $0.8$ (Canopy),   & $0.048$ (Canopy)& &  \\
&Plage&$>5.0$ (Plage)&$0.059$ (Plage)& & Figure \ref{fig:iris_obs} \\ 
&Umbra&$0.85$ (Umbra)&$0.025$ (Umbra)& & Umbra is single peaked\\
&&&&& \\
{\tt nw072100} & QS & $0.6$--$1.5$ & $0.03$ &  $100$ & Figure~\ref{fig:qs_profiles} \\ 
{\tt qs072100} & QS & $2.5$--$3.5$ & $0.025$ & $100$ &\ktwor\ stronger\\ 
&&&&& Figure~\ref{fig:qs_profiles} \\
&&&&& \\
{\tt QS GOL} & QS & $0.8$ & $0.04$ & 5 & \ktwor\ stronger\\
&&&&& Figures~\ref{fig:qs_tgstruct_gol},\ref{fig:sim4km_profiles}\\
{\tt QS GOL, NEQ(H)} & QS & $1.2$ & 0.04 & 5 & \ktwor\ stronger\\
&&&&& Figures~\ref{fig:qs_tgstruct_gol},\ref{fig:sim4km_profiles}\\
&&&&& \\
{\tt Spicule $2.5$D} & Spicule & $1.1$ (nGOL), & 0.04 &14 &\ktwov\ strong/single peaked\\
{\tt GOL}&&$1.4$ & 0.04 &14& \\
{\tt GOL, NEQ(H,He)}&&$3.2$&$0.05$& $14$ & Figures~\ref{fig:spicule_tgstruct_gol},\ref{fig:spicule_profiles_gol_hhe}\\
&&&&& \\
{\tt pl072100} & Plage & $7$--$20$ & $0.025$ & $100$ &Figure~\ref{fig:plage_panels} \\
&&&&& Plage single peaked \\
{\tt nw072100} & FE & 1 (``QS'') & $0.05$ (``QS'')  & 100& ``AR/Plage'' single peaked \\
&&10 (``AR'') &$0.06$ (``AR'')& &Figure~\ref{fig:postfe_profiles} \\
\end{tabular}
\caption{\label{tab:mgii_summary} Summary of \mgii\ properties for observed and synthetic profiles. The following abbreviations are used: Quiet Sun (QS), Active Region (AR), Flux emergence (FE), internetwork (IN), network (NW) non-equilibrium NEQ, non-GOL (nGOL). }
\end{table*}

The importance of the magnetic field strength and structure is also apparent in the models featuring flux emergence. In the most active portion of the post emergence {\tt nw072100} model (outlined in blue in Figures~\ref{fig:fe_profiles} and \ref{fig:mg_prof_atmos}) hot coronal loops reach temperatures of $>5$~MK. This leads to high coronal pressures and a highly compressed chromosphere. This chromosphere, while geometrically foreshortened, is dense and hot in its upper portion (in similarity with the plage model {\tt pl072100}). The heating driving the high chromopheric and coronal temperatures comes both from high angle reconnection of emerging and pre-existing ambient field lines and from the small angle reconnection due to the braiding of already present field lines (e.g. as in Bose et al. 2022). Note that while \mgii~k line intensities are quite high, higher than those observed in plage, and while the profiles are single peaked, their average width is significantly smaller than that observed.

We note that the column mass, $m_c$ can be a good proxy for optical depth, and that $\log_{10}(m_c)=-2.2$ (kg/m$^2$) has a rough correspondance with the height of \kthree\ emission. An overview of the magnetic field at coronal heights, the column mass height, the Joule heating per unit mass in the chromosphere (1.5~Mm above the photosphere), and the emission in the \feix\ 17.1~nm spectral line showing sites of strong coronal heating and temperatures are displayed in Figure~\ref{fig:mg_prof_atmos}. The figure also contains plots of average and individual line profiles as well as the chromospheric structure of  the temperature, vertical velocity, and particle densities as a function of height above the photosphere. The average vertical fields in the larger green, red and blue regions of interest are $\uBz = 9, 37, 150$~Gauss respectively and the chromospheric heating is much stronger in the central blue region than in relatively more quiet green and red areas. However, note the strong shock structures seen in second, green, row and also the large density scale height in both the red and green rows. The \feix\ image implies that both of these locations are covered by cool canopy-like material. The emission from the central, blue, region is extremely bright and is formed over a compressed dense chromosphere with the transition region to the corona placed only slightly above 1~Mm above the photosphere. 

 The role of the global magnetic field is illustrated by considering regions far away from the strongest fields.  In Figure~\ref{fig:postfe_profiles} profiles from late in the {\tt nw072100} model are shown. At this stage of the simulation, profiles that are nearly identical to those found in the quiet Sun and in the average spectra of a small active region are found. While the average vertical field strength $\uBz$ at this stage of the simulation is high, higher than that measured in the quiet Sun, we find that the average photospheric field in the approximately $10\times 10$~Mm regions (outlined in green and red in Figure~\ref{fig:fe_profiles}), directly below the chromosphere forming the \mgii\ lines is only 10 -- 20~Gauss: It is the larger scale field forming longer loops above the regions of interest that plays the leading role in forming chromospheric structure by bringing cold material to great heights and thereafter holding it aloft. A key question of course is whether such a scenario represents the quiet Sun environment on the Sun well, even if the spectral line properties are improved. After all, the sequence of events in {\tt nw072100} involves prior flux emergence on very large scales that is key to bring cold material to great heights, subsequent weakening and dispersal of the magnetic field, while holding the previously injected cold material at great heights. Such large-scale flux emergence is not the typical cause for quiet Sun magnetism, which is thought to be sustained through the continual emergence of much smaller-scale ephemeral regions as well as, to a lesser extent, decaying active regions. Perhaps the relative success of {\tt nw072100} points instead to the key role of mass loading into the chromosphere, which likely is not properly captured by current simulations.
 
\section{Conclusions and discussion}\label{sec:con}

We have found that key contributors to the observed widths are the chromospheric heating, mass loading and spatial extent, while the \mgii\ intensities are strongly coupled to both chromospheric and coronal heating. While these conclusions were already clear from the semi-empirical modeling of plage in \citet{2015ApJ...809L..30C}, the forward models presented here show possible physical conditions and processes required in order to form chromospheres that can reproduce the observations. 

We present several quiet Sun models, listed in Table~\ref{tab:mgii_summary}, those with coarse resolution have widths of 0.025~nm, while both the non-GOL and GOL models with high 5~km resolution are substantially broader at 0.04~nm. This is a similar sensitivity to resolution as in the case for the \ion{Ca}{2}~854.2~nm line shown in Figure~\ref{fig:caii_resolution}. However, we note that there still is a significant difference between observed quiet Sun widths, which are of order 0.05~nm for internetwork and 0.056~nm for network.

It is first when we introduce more mass at greater heights into the chromosphere than what is typically found in semi-empirical models, such as the 2~Mm high VAL3C semi-empirical model \citep{1981ApJS...45..635V} or the {\tt qs072100} model presented here, that we find widths approaching those observed. 

The results of models featuring emerging flux regions or spicules, where mass is carried or thrown into the upper chromosphere and lower corona, show a much improved correspondence with \iris\ observations. This is true both in terms of line intensities and line widths. The magnetic field  is the key player in these phenomena, underscoring the importance of capturing both the magnetic field strength and topology in simulations. On the other hand, the plage model and the more active parts of the flux emergence simulation still produce \mgii\ profiles with widths smaller than those observed, indicating that we are likely missing something important in our understanding of the more active Sun's and unipolar plage's geometry, dynamics, and heating processes.  

It seems clear that turbulent motions alone are not likely to be the sole root cause of the observed \mgii\ core line widths. This is because the observed discrepancy would require very large values of turbulence that appear to be incompatible with current observations from other lines, e.g., \ion{O}{1} 1355\AA\ \citep{2015ApJ...809L..30C}. However, it should be noted that it is not fully clear whether the line formation region of this optically thin line is very similar to that of \ion{Mg}{2} (as suggested by \citet{2015ApJ...813...34L} for quiet Sun simulations) for all solar regions. 

It is interesting to note that the good correspondence found in quieter regions, that have undergone significant mass loading through large-scale flux emergence, is possible even with the relatively coarse resolution of 100~km. This suggests that the heating of injected mass associated with the field, and hence opacity broadening, is a key agent of the large \mgii\ widths. The question then is which process dominates this mass supply, with both spicules and flux emergence candidate processes considered in this paper. Neither of these processes is fully captured with current models. The spicule models we have shown include GOL and non-equilibrium ionization, but are limited to 2D. The current simulations provide a better match with observed spectral line properties but are not sufficient. It remains to be seen whether 3D models can further reduce the discrepancies with observations. Similarly, flux emergence models typically include emergence on very large spatial scales that are a good fraction of an active region size. Some of these models during late stages similarly provide a much better match with observed spectra, with discrepancies much reduced. While such large-scale emergence clearly plays a dominant role in the formation of active regions, it is not clear whether the scenario involved in the large-scale emergence applies to the formation of quiet Sun for which smaller-scale emergence is thought to play a key role. Recent results from sufficiently high resolution and deep numerical models \citep{2014ApJ...789..132R,2018ApJ...859..161R} predict that a local dynamo will be active, even in the quiet Sun away from any contribution from the global dynamo, producing fields of order $\uBz\approx 60$~Gauss. This may be sufficient to cause ``continual'' flux emergence, but whether the field will be strong enough to significantly perturb  chromospheric structure remains to be seen. It is also not clear whether such models include sufficient numbers of medium-scale ephemeral regions which are known to affect the chromosphere \citep{2021ApJ...911...41G} and which are thought to play, through continual emergence, a dominant role in supplying the quiet Sun network. 

It is also clear from our work that very strong magnetic field regions such as plage continue to represent a challenge in terms of reproducing the observed properties of \ion{Mg}{2}. It is unclear whether mass loading alone can resolve this issue. Our results indicate that non-equilibrium ionization and non-MHD effects such as GOL play an important role in reducing discrepancies with observations. However these effects are computationally expensive and have not been studied for all different types of simulations. In addition, it is still unclear whether these effects can further reduce remaining discrepancies, for example, at higher field strengths, which are known to occur in plage. Another unknown aspect is whether other multi-fluid effects, such as the Thermal Farley Buneman instability, which have not been studied under realistic chromospheric conditions but are thought to potentially play a role in the chromosphere \citep{2020ApJ...891L...9O, 2022arXiv221103644E}, can significantly reduce discrepancies. 

Our results provide a path forward for further studies focusing on understanding the formation of \ion{Mg}{2}. The results of higher numerical resolution simulations indicate that a higher resolution appears to significantly reduce discrepancies with observations, suggesting that the numerical approach plays a key role. This likely goes beyond the direct effects of higher velocities on the broadening of the line, but also includes the significant increase in heating and mass loading as a result of higher resolution. Furthermore, having a large spatial extent, allowing field topologies at many scales, is clearly important. The challenge then is to produce numerical simulations that combine many of these different effects in order to determine whether the combination of these various impacts leads to a full explanation of the average \ion{Mg}{2} profiles. Beyond that, it is also clear that much can be learned from studying the spatial distribution and temporal evolution of the simulated profiles, as well as from investigating where in the simulations the discrepancies are not present. That will all be part of future work.



\acknowledgements{\longacknowledgment} 

\bibliographystyle{aasjournal}
\bibliography{aamnemonic,mgii_refs}

\end{document}